\documentclass[showpacs,preprintnumbers,amsmath,amssymb]{revtex4}

\usepackage{graphicx}
\usepackage{dcolumn}
\usepackage{bm}

\begin{document}

\title{Spin relaxation in a complex environment}

\author{Massimiliano Esposito}
\author{Pierre Gaspard}
\affiliation{Center for Nonlinear Phenomena and Complex Systems,\\
Universit\'e Libre de Bruxelles, Code Postal 231, Campus Plaine, B-1050
Brussels, Belgium.}

\date{\today}
\pacs{05.50.+q, 03.65.Yz}

\begin{abstract}
We report the study of a model of
a two-level system interacting in a nondiagonal way with a
complex environment described by Gaussian
orthogonal random matrices (GORM). The effect of the
interaction on the total spectrum and its consequences on the
dynamics of the two-level system is analyzed. We show the existence of a
critical value of the interaction, depending on the mean level
spacing of the environment, above which the dynamics is self-averaging
and closely obey a master equation for the time evolution of the
observables of the two-level system.  Analytic results are also obtained
in the strong coupling regimes.  We finally study the equilibrium values 
of the two-level system population and show under which condition it
thermalizes to the environment temperature.
\end{abstract}

\maketitle
\section{Introduction}

The nonequilibrium statistical mechanics of small quantum systems has
become a topic of fundamental importance for nanosciences.  It is indeed
very important to understand what the minimum sizes and conditions are under which
a quantum system can display a relaxation to an
equilibrium state.  Isolated and finite quantum systems have a discrete energy
spectrum, which has for consequence that all the observables present
almost periodic recurrences on long time scales.  Nevertheless, their early
time evolution may still present relaxation types of behavior that is important
to study.  Although tools have been developed in nonequilibrium statistical mechanics 
to describe such relaxations to a state of equilibrium by master and kinetic equations, 
the conditions of validity of these equations remain little known.

It is the purpose of this paper to contribute to the clarification of these questions
of validity of the kinetic description by studying of a simple model
of a two-level system or spin coupled to a complex environment
described by random matrices.  Indeed, work done during the last decade
has shown that the Hamiltonian of typical quantum systems presents
properties of random matrices on their small energy scales. Here,
we consider the Hamiltonian of the environment as well as the operator
of coupling between the spin and the environment to be given by
random matrices taken in a statistical ensemble of Gaussian orthogonal 
random matrices. This defines a model in which many results can be
obtained analytically.  

Similar models using Gaussian orthogonal random matrices 
\cite{Lebowitz,Rice} or using banded random matrices 
\cite{Pereyra1,Pereyra2,Lutz,Cohen1,Cohen2} have been studied. 
In \cite{Lutz}, it has been shown that random matrices used as 
environment coupling operators have a universal feature.

The model we here consider differs from the spin-boson
model by the density of states of the environment. Instead of
monotonously increasing with energy as in the spin-boson model,
the density of states of the environment obeys Wigner's semicircle law
in our model and is thus limited to an interval of energy with a maximum
density in between. We notice that such densities of states appear
in systems where a spin is coupled to other (possibly dissimilar) spins
as, for instance, in NMR, in which case the density of states of the other spins
forming the environment also present a maximum instead of a monotonous
increase with energy.  Our model may therefore constitute a simplification
of such kinds of interacting spin systems.  Our main purpose is to understand
the conditions under which a kinetic description can be used in order
to understand the relaxation of the spin under the effect of the coupling
with the rest of the system, which we refer to as a complex environment.

The plan of the paper is the following. The model is presented in
Sec. \ref{model}.  The properties of the spectrum are described in
Sec. \ref{spectrum}. The relaxation in the time evolution
of the spin is studied in Sec. \ref{time}.  The very long time
behavior and the approach to the equilibrium is discussed in Sec. \ref{equilibrium}.
Conclusions are drawn in Sec. \ref{conclusions}.

\section{The model}
\label{model}

We are interested in the study of a total system composed of a simple
system (with a few discrete levels) interacting with a complex environment
(with many levels). We consider a two-level system as a prototype for the simple system.

For this kind of total system, the time-dependent Schr\"{o}dinger equation 
is of the following type:\\
\begin{eqnarray}
i \hbar \frac{d \vert \Psi (\tilde{t}) \rangle}{d\tilde{t}} &=&
\hat{\tilde{H}}_{\rm tot} \vert \Psi (\tilde{t}) \rangle \nonumber \\
&=& \left( \frac{\tilde{\Delta}}{2} \hat{\sigma}_{z} +
\hat{\tilde{H}}_{B} + \tilde{\lambda} \hat{\sigma}_x
\hat{\tilde{B}} \right)  \vert \Psi (\tilde{t}) \rangle ,
\label{S-s-GOE}
\end{eqnarray}
where \\ $\bullet$ $\hat{\sigma}_{x}$, $\hat{\sigma}_{y}$, and $\hat{\sigma}_{z}$ are the $2
\times 2$ Pauli matrices,\\ $\bullet$ $\frac{\tilde{\Delta}}{2}
\hat{\sigma}_{z}$ is the Hamiltonian of the two-level system,\\ $\bullet$
$\tilde{\Delta}$ is the energy spacing between the two levels of
the system,\\ $\bullet$ $\hat{\tilde{H}}_{B}$ is the Hamiltonian of the
environment,\\ $\bullet$ $\hat{\sigma}_x$ is the coupling operator of
the system,\\ $\bullet$ $\hat{\tilde{B}}$ is the coupling operator of
the environment,\\ $\bullet$ $\tilde{\lambda}$ is the coupling parameter
between the system and the environment. \\

The well-known \textit{spin-boson model} \cite{Suarez,Leggett,GaspRed}
is a particular case of the total system where $\hat{\tilde{H}}_{B}$ corresponds to an
infinite harmonic oscillator lattice and $\hat{\tilde{B}}$ is
linear in the degree of freedom of the environment. 

Here, we want to define a new model, the \textit{spin-GORM model}, 
also described by the Hamiltonian
(\ref{S-s-GOE}) and for which $\hat{\tilde{H}}_{B}$ and
$\hat{\tilde{B}}$ are Gaussian orthogonal random matrices (GORM)
(see appendix A for some basic property on GORM).

Let us discuss now the spin-GORM model in more detail.

As we said, we want to model a two-level system that interacts with an
environment that has a complex dynamics. Here, complex is
used in a generic way. The complexity can come, for example, from
the fact that the corresponding classical system is chaotic like
in a quantum billiard or for the hydrogen atom in a strong
magnetic field \cite{Guhr,BGS}. It can also come from large
coupling in an interacting many-body
system as in nuclear physics \cite{Guhr} or in
interacting fermion systems such as quantum computers
\cite{Guhr,Shepe}. Wigner in $1960$ \cite{Mehta,Porter,Brody} was the first
to develop random-matrix theory for the purpose of modeling spectral
fluctuations of complex quantum systems containing many states
interacting with each other. This tool has now become very common
in many fields from nuclear physics to quantum chaos. This is the
reason why we consider random matrices to characterize the
complexity of the environment operators.

The environment operators of the spin-GORM model,
$\hat{\tilde{H}}_B$ and $\hat{\tilde{B}}$, are defined by
\begin{eqnarray}
\hat{\tilde{H}}_B &=& \sigma_{ND}^{\hat{\tilde{H}}_B} \hat{X}
\nonumber \\ \hat{\tilde{B}} &=& \sigma_{ND}^{\hat{\tilde{B}}}
\hat{X}^{'}, \label{bof1}
\end{eqnarray}
where $\hat{X}$ and $\hat{X}^{'}$ are two
different $\frac{N}{2} \times \frac{N}{2}$
Gaussian orthogonal random matrices with mean zero.
Their nondiagonal (diagonal) elements have standard deviation $\sigma_{ND}^{\hat{\tilde{X}}}= 1$ 
($\sigma_{D}^{\hat{\tilde{X}}} = \sqrt{2}$). $\hat{X}$ and $\hat{X}^{'}$ are two
different realizations of the same random matrix ensemble and have
therefore the same statistical properties. 
$\sigma_{ND}^{\hat{\tilde{H}}_B}$ and
$\sigma_{ND}^{\hat{\tilde{B}}}$ are the standard deviations of the
nondiagonal elements of $\hat{\tilde{H}}_B$ and $\hat{\tilde{B}}$,
respectively. For these random matrices, the width of their
averaged smoothed density of state is given by
\begin{eqnarray}
\mathcal{D}\!\tilde{H}_B = \sigma_{ND}^{\hat{\tilde{H}}_B} \sqrt{8 N}, \nonumber \\
\mathcal{D}\!\tilde{B} = \sigma_{ND}^{\hat{\tilde{B}}} \sqrt{8 N} 
\end{eqnarray}
(see Appendix A).

It is interesting to define the model in such a way that, when $N$
is increased, the averaged smoothed density of state of the
environment increases without changing its width
$\mathcal{D}\!\tilde{H}_B$. The width can be fixed to unity. This
is equivalent to fixing the characteristic time scale of the
environment. For doing this, it is necessary to rescale the
parameters as follows:
\begin{equation}
\left\lbrace
\begin{array}{l}
\alpha=\sigma_{ND}^{\hat{\tilde{H}}_B} \sqrt{8 N} \; , \\
\\
t=\alpha \tilde{t} \nonumber \; ,\\
\\
\Delta = \frac{\tilde{\Delta}}{\sigma_{ND}^{\hat{\tilde{H}}_B} \sqrt{8 N}} \; ,\\
\\
\lambda=\tilde{\lambda} \frac{\sigma_{ND}^{\hat{\tilde{B}}}}{\sigma_{ND}^{\hat{\tilde{H}}_B}} \; ,\\
\\
N=N .
\end{array}
\right.
\end{equation}
\\
The time-dependent Schrodinger equation of the spin-GORM model
becomes
\begin{equation}
i \hbar \frac{d \vert \Psi (t) \rangle}{dt} = \hat{H}_{\rm tot}
\vert \Psi (t) \rangle ,
\end{equation}
with the rescaled total Hamiltonian
\begin{eqnarray}
\hat{H}_{\rm tot} &=&  \hat{H}_{S}  +  \hat{H}_{B} +  \lambda \hat{\sigma}_x \hat{B} \nonumber \\
 &=&  \frac{\Delta}{2} \hat{\sigma}_{z}  +  \frac{1}{\sqrt{8N}} \hat{X}  +
 \lambda  \hat{\sigma}_x  \frac{1}{\sqrt{8N}} \hat{X}^{'} . \label{hamiltonien}
\end{eqnarray}
As announced, we have now $\mathcal{D}\!H_B=\mathcal{D}\!B=1$.

In the following, without loss of generality, $\alpha$ will always be
taken equal to unity. Notice that, to model an environment with a
quasicontinuous spectrum, the random matrix must be very large ($N \to
\infty$).

In order to get ensemble averaged results, one has to
perform averages over the different results obtained for each
realization of Eq. (\ref{hamiltonien}). When we use finite
ensemble averages, the number of members of the ensemble average
will be denoted by $\chi$.

We see that the Hamiltonian (\ref{hamiltonien}) is characterized
by three different parameters: $\Delta$, $\lambda$, and $N$. We
define three different parameter domains in the reduced parameter
space corresponding to a fixed $N$ in order to facilitate the
following discussion. These three regimes are represented in
Fig. \ref{regimeschema}: domain A with $1 > \lambda, \Delta$; 
domain B with $\Delta > 1,\lambda$; domain C with $\lambda > 1,\Delta$.

\section{The spectrum}
\label{spectrum}

In this section we study the spectrum of the complete system for the different
values of the parameters. This study is important in order to
understand the different dynamical behaviors that we 
encounter in the model.

Let us begin defining the notations in the simple case where there
is no coupling between the two parts of the total system ($\lambda
\to 0$). The isolated system has two levels separated by the energy $\Delta$:
\begin{equation}
\hat{H}_{S} \vert s \rangle = s \frac{\Delta}{2} \vert s \rangle,
\end{equation}
where $s=\pm 1$. The environment has the standard spectrum of a
GORM
\begin{equation}
\hat{H}_{B} \vert b \rangle = E_{b}^{B} \vert b \rangle,
\label{VPenviron}
\end{equation}
where $b=1,2,...,N/2$. The Hamiltonian of the total system without
interaction between the system and the environment is thus
\begin{equation}
\hat{H}_{0}=\hat{H}_{S}+\hat{H}_{B},
\end{equation}
and the spectrum is therefore given by
\begin{equation}
\hat{H}_{0} \vert n \rangle = E^{0}_{n} \vert n \rangle,
\label{eqVPHo}
\end{equation}
with $ n=1,2,...,N$ and 
\begin{equation}
E^{0}_{n} = s \frac{\Delta}{2} + E_{b}^{B}. \label{eqVPHobis}
\end{equation}
The eigenvectors are tensorial products of both the system and 
environment eigenvectors:
\begin{equation}
\vert n \rangle = \vert s \rangle \otimes \vert b \rangle.
\end{equation}

Let us now define the notations in the opposite simple situation
where the coupling term is so large that the Hamiltonian of the
system and of the environment can  both be neglected ($\lambda \to
\infty$). Using the unitary matrix $\hat{U}$ acting only on the
system degrees of freedom
\begin{equation*}
\hat{U}=
\begin{bmatrix}
\frac{1}{\sqrt{2}} & \frac{1}{\sqrt{2}} \\ \frac{1}{\sqrt{2}} &
-\frac{1}{\sqrt{2}} \\
\end{bmatrix},
\end{equation*}
the total Hamiltonian becomes
\begin{eqnarray}
\tilde{\hat{H}}_{0} = \lambda \hat{\sigma}_z \hat{B}.
\end{eqnarray}
$E_{\kappa \eta}$ and $\vert \kappa \eta \rangle = \vert \kappa
\rangle \otimes \vert \eta \rangle$ are, respectively, the
eigenvalues and eigenvectors of the Hamiltonian
\begin{equation}
\tilde{\hat{H}}_{0} \vert \kappa \eta \rangle = \lambda E_{\kappa
\eta} \vert \kappa \eta \rangle = \lambda \kappa E_{\eta} \vert
\kappa \eta \rangle ,
\label{eqVPHoinv}
\end{equation}
where $\eta=1,...,\frac{N}{2}$ and $\kappa=\pm1$.\\

After having defined the notation in the two extreme cases
$\lambda \to 0$ and $\lambda \to \infty$, we will start the study
of the spectrum with interaction $\lambda \neq 0$.

The total spectrum is given by the eigenvalues $\lbrace E_{\alpha} \rbrace$ that are solutions
of the eigenvalue problem:
\begin{equation}
\hat{H}_{\rm tot} \vert \alpha \rangle = E_{\alpha} \vert \alpha
\rangle, \label{eqVPtot}
\end{equation}
where $\alpha=1,2,...,N$. It is very difficult to obtain
analytical results for this problem. We will therefore study the
total spectra using a method of numerical diagonalization of
the total Hamiltonian.

\subsection{Smoothed density of states}

In order to have a quantitative understanding of the global aspect
of the spectrum (on large energy scales), we will study the total
perturbed averaged smoothed density of states.

The environment-averaged smoothed density of states obeys the
semicircle Wigner law [see Eq. (\ref{semi-circ}) in
Appendix A]:
\begin{eqnarray}
n^w(\epsilon) &=& \frac{4N}{\pi} \sqrt{(\frac{1}{2})^2-\epsilon^2}
\ \ \mbox{if} \ \ \vert \epsilon \vert < \frac{1}{2} \nonumber \\ &=& 0 \
\ \mbox{if} \ \ \vert \epsilon \vert \geq \frac{1}{2},
\end{eqnarray}
where $\epsilon$ is the continuous variable corresponding to the
environment energy $E_{b}^{B}$. 

\begin{figure}[h]
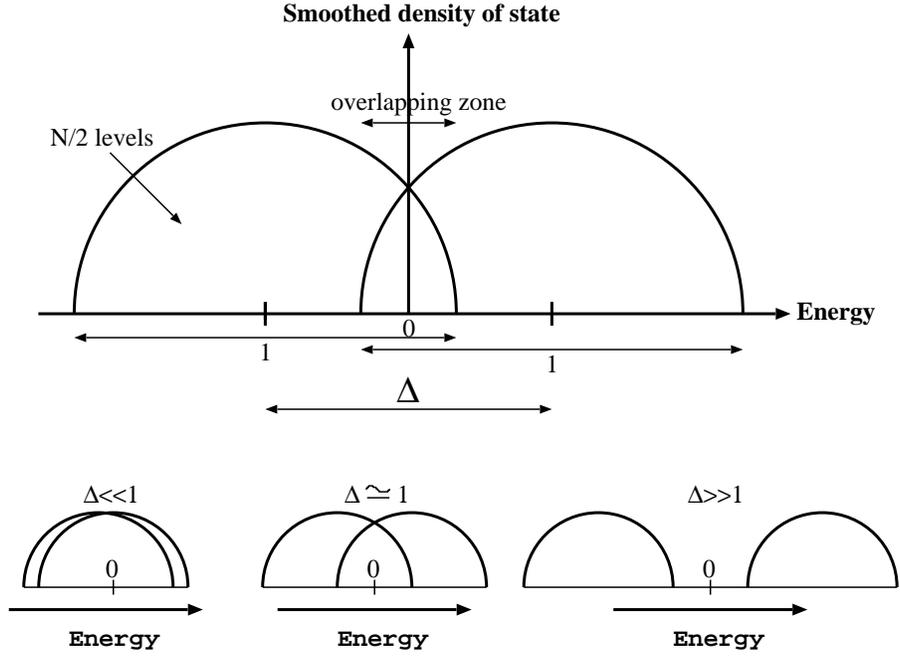

\centerline{\rotatebox{0}{\scalebox{1}{\includegraphics{fig2.eps}}}}
\vspace*{0.8cm}
\centerline{\rotatebox{0}{\scalebox{1}{\includegraphics{fig3.eps}}}}
\caption{Smoothed densities of states $n^w(\varepsilon-\frac{\Delta}{2})$ and
$n^w(\varepsilon+\frac{\Delta}{2})$ 
for different values of $\Delta$. The total smoothed averaged
density of states of the nonperturbed spectrum is obtained by the
sum of them (see Eq. (\ref{doubledistrib})).} \label{troisdelta}
\end{figure}

\begin{figure}[h]
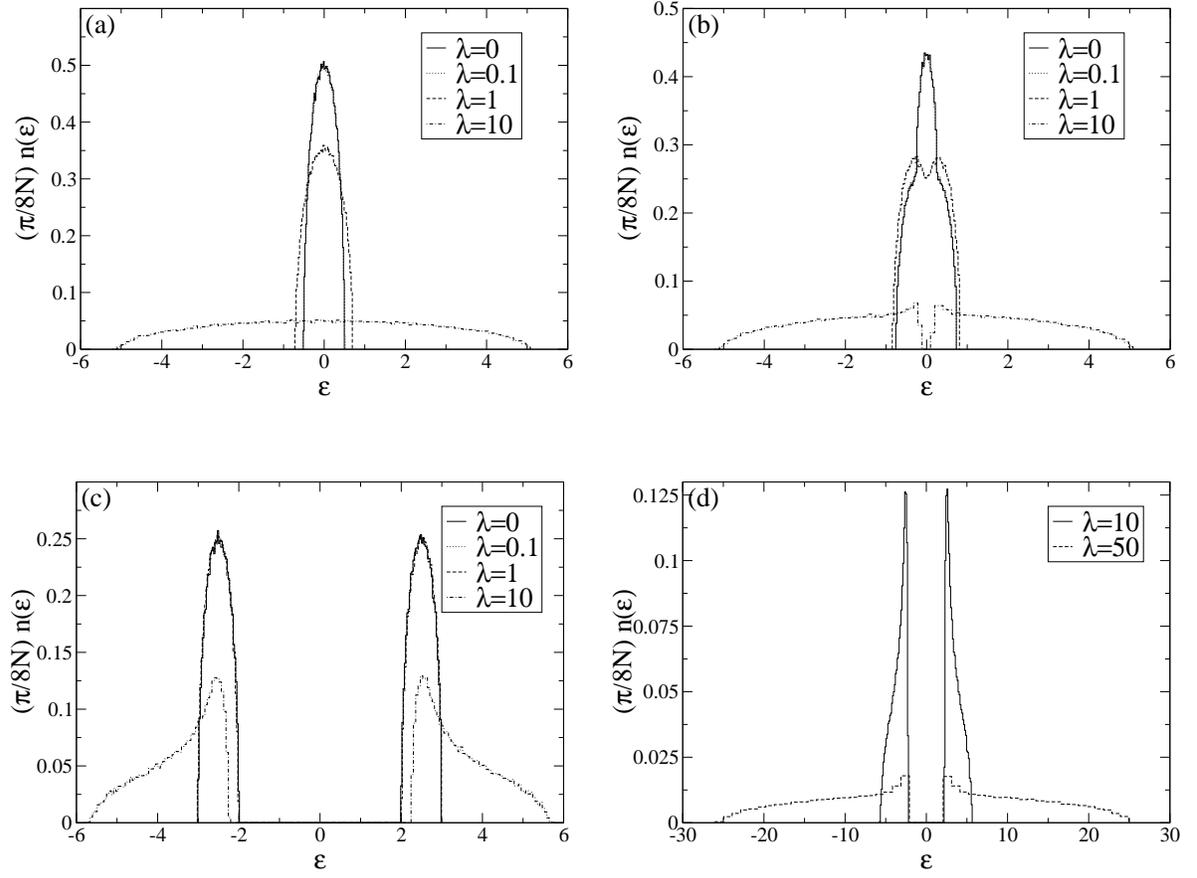

\centering
\begin{tabular}{c@{\hspace{0.5cm}}c}
\vspace*{1cm}
\rotatebox{0}{\scalebox{0.3}{\includegraphics{fig4.eps}}} &
\rotatebox{0}{\scalebox{0.3}{\includegraphics{fig5.eps}}} \\
\rotatebox{0}{\scalebox{0.3}{\includegraphics{fig6.eps}}} &
\rotatebox{0}{\scalebox{0.3}{\includegraphics{fig7.eps}}} \\
\end{tabular}
\caption{Total smoothed averaged
density of states obtained numerically for different values of
$\Delta$ and $\lambda$. (a) corresponds to $\Delta=0.01$, (b)
to $\Delta=0.5$, and (c) and (d) to $\Delta=5$. For all of them $N=500$
and $\chi=50$. Notice that the $\lambda=0$ and the $\lambda=0.1$ 
curves are not distinguishable.} \label{smoothedplot}
\end{figure}

Therefore, when $\lambda=0$, the total averaged smoothed density
of states is the sum of the two environment semicircle 
densities of states which correspond to both states of the
two-level system [see Eq. (\ref{eqVPHo})]:
\begin{eqnarray}
n(\varepsilon) &=&
n^w(\varepsilon-\frac{\Delta}{2})+n^w(\varepsilon+\frac{\Delta}{2})
\nonumber
\\ &=& \frac{4N}{\pi}
\sqrt{(\frac{1}{2})^2-(\varepsilon-\frac{\Delta}{2})^2} +
\frac{4N}{\pi}
\sqrt{(\frac{1}{2})^2-(\varepsilon+\frac{\Delta}{2})^2} \nonumber
\\ && \mbox{if} \ \ (\frac{1}{2}-\frac{\Delta}{2}) < \vert \varepsilon
\vert < (\frac{1}{2}+\frac{\Delta}{2}) \nonumber \\ &=& 0 \ \
\rm{elsewhere} \label{doubledistrib}
\end{eqnarray}
where $\varepsilon$ is the continuous variable corresponding to
the total energy $E^{0}_{n}$. 
The semicircle densities
$n^w(\varepsilon-\frac{\Delta}{2})$ and
$n^w(\varepsilon+\frac{\Delta}{2})$ are schematically depicted in Fig. \ref{troisdelta} 
for different values of $\Delta$. The numerical density of states
of the total system
corresponding to $\lambda=0$ is depicted in Fig. \ref{smoothedplot}.

When $\lambda \to \infty$ (meaning that the coupling term becomes
dominant in the Hamiltonian), the averaged smoothed density of
states of the total system (see Eq.(\ref{eqVPHoinv})) is
given by
\begin{eqnarray}
n(\varepsilon) = n^w(\frac{\varepsilon}{\lambda}) + n^w(- \frac{
\varepsilon}{\lambda}) &=& \frac{8N}{\lambda \pi}
\sqrt{(\frac{\lambda}{2})^2-(\varepsilon)^2} \ \ \mbox{if} \ \ \vert
\varepsilon \vert < \frac{\lambda}{2} \nonumber \\ &=& 0 \ \ \mbox{if} \
\ \vert \varepsilon \vert \geq \frac{\lambda}{2}, \label{1surleg0}
\end{eqnarray}
where $\varepsilon$ is the continuous variable corresponding to
the total energy $E_{\kappa \eta}$. This result can be observed
in Fig.\ref{smoothedplot}(a) for $\lambda=10$ (because
$\Delta,1 \ll \lambda$).

When $\lambda \neq 0$, the total averaged smoothed
density of states is also plotted in Fig. \ref{smoothedplot}. The
main observation is that there is a broadening of the complete
spectrum when one increases $\lambda$. In Fig. \ref{smoothedplot}(a), we see that the
averaged smoothed density of states changes in a smooth way from
(\ref{doubledistrib}) to (\ref{1surleg0}). But in Fig. \ref{smoothedplot}(b) and much
more in Figs. \ref{smoothedplot}(c) and (d), the two semicircle densities
$n^w(\varepsilon-\frac{\Delta}{2})$ and
$n^w(\varepsilon+\frac{\Delta}{2})$ seem to repel each other as
$\lambda$ increases. This is due to the fact that the levels of a
given semicircle density don't interact with each other but only interact with 
the levels of the other semicircle density. This is a consequence of the nondiagonal 
form of the coupling. Therefore, having in mind the perturbative expression of
the energies [see Eq. (\ref{pertcas2}) of Appendix C], one understands
that when $\Delta$ is nonzero, the eigenvalues that are repelling
each other with the most efficiency are the ones closest to the
center of the total spectrum.

\subsection{Eigenvalue diagrams}

The global effect of the increase of $\lambda$ on the eigenvalues
has been studied with the average smoothed density of states.
But in order to have an idea of what happens on a finer energy
scale inside the spectrum, it is interesting to individually follow
each eigenvalue $E_{\alpha}$ as a function of $\lambda$ on an
eigenvalue diagram.

\begin{figure}[h]
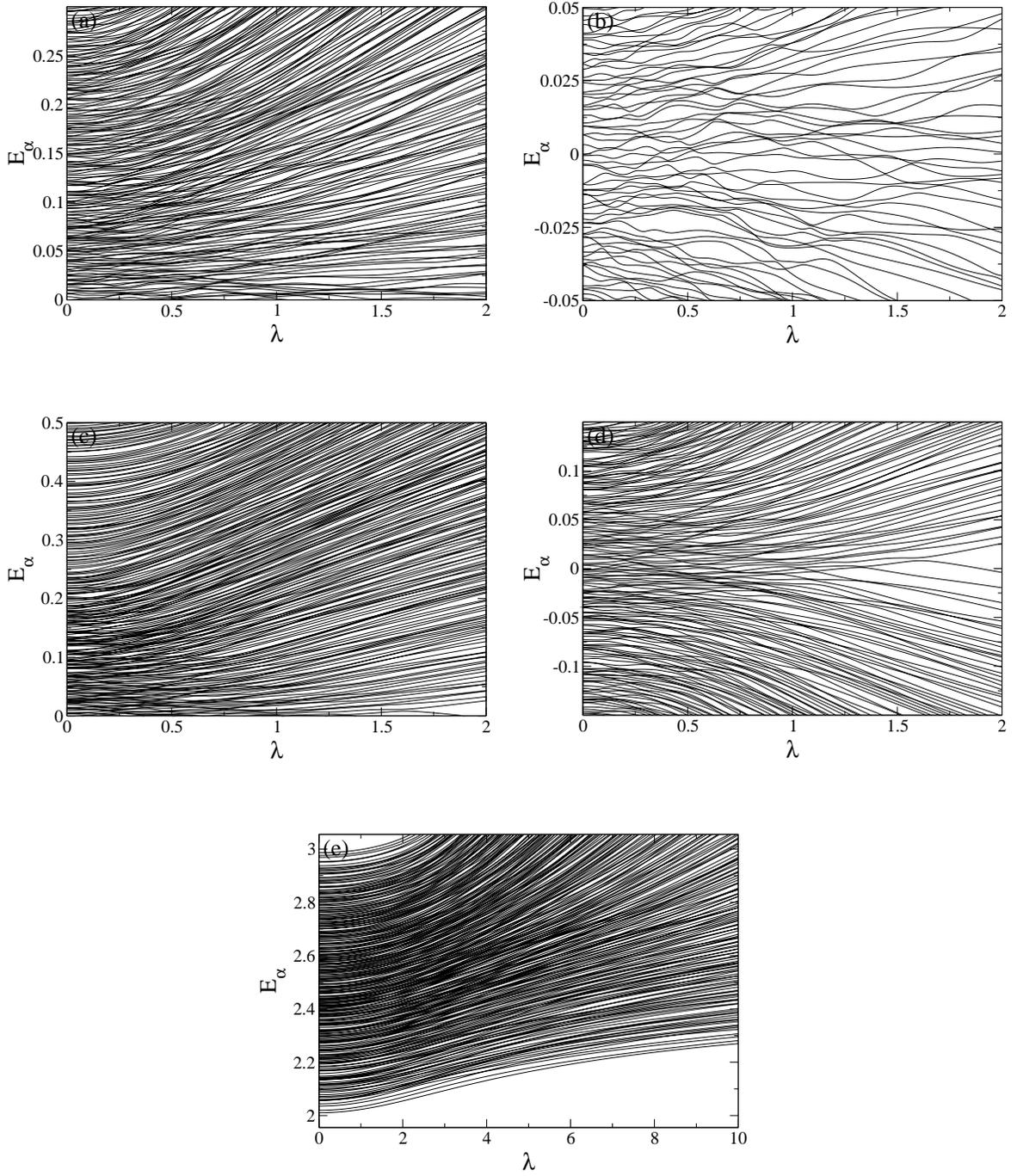

\centering
\begin{tabular}{c@{\hspace{0.5cm}}c}
\vspace*{1cm}
\rotatebox{0}{\scalebox{0.3}{\includegraphics{fig8.eps}}}
&
\rotatebox{0}{\scalebox{0.3}{\includegraphics{fig9.eps}}}
\\
\vspace*{1cm}
\rotatebox{0}{\scalebox{0.3}{\includegraphics{fig10.eps}}}
&
\rotatebox{0}{\scalebox{0.3}{\includegraphics{fig11.eps}}}
\\
\end{tabular}
\centering
\rotatebox{0}{\scalebox{0.3}{\includegraphics{fig12.eps}}}
\caption{Different parts of the eigenvalue diagrams with $N=500$, corresponding
to (a) and (b) $\Delta=0.01$, (c) and (d) $\Delta=0.5$, and (e) $\Delta=5$. They represent the
eigenvalues of the total Hamiltonian (\ref{eqVPtot}) as a function
of $\lambda$.} \label{croisementd=0.01etd=0.5}
\end{figure}

The first thing to note (see
Fig. \ref{croisementd=0.01etd=0.5}) is that the increase of the
coupling induces a repulsion between the eigenvalues. This has, as a
result the broadening of the total spectrum as we already noticed
on the smoothed averaged density of states. If one looks closer
inside the fine structure of the eigenvalue spectrum, we see that there
is no crossing between the eigenvalues. This is a consequence of the
fact that there is no symmetry in the total system. Therefore, the
non-crossing rule is always working. Each time two
eigenvalues come close to each other, they repel each other and
create an avoided crossing. One can notice that there is a
large number of avoided crossings inside the region where the two
semicircle $n^w(\varepsilon-\frac{\Delta}{2})$ and
$n^w(\varepsilon+\frac{\Delta}{2})$ overlap (see
Fig. \ref{troisdelta} in order to visualize the overlapping
zone that extends from $\frac{\Delta}{2}-1$ to
$1-\frac{\Delta}{2}$). But outside this overlapping zone, there are
very weak avoided crossings (and of course no crossing) and all
the eigenvalues appear to move in a regular and smooth way. The regions seen
in Figs. \ref{croisementd=0.01etd=0.5}(a), (b) and (d) are inside 
the overlapping zone and the one in Fig. \ref{croisementd=0.01etd=0.5}(e) is outside. 
In Fig. \ref{croisementd=0.01etd=0.5}(c), the lower part of the
diagram is inside the overlapping zone and the upper part is outside. This
phenomenon is due to the fact that the eigenvalues of a given
semicircle density do not interact with the eigenvalues of their own
semicircle density but only with those of the other
semicircle density. It is the consequence of the nondiagonal nature of
the coupling in the spin degrees of freedom. When the coupling
becomes large enough ($\lambda
> 1$), the avoided crossings also disappear inside the overlapping zone.
Around $\lambda=1$ and inside the overlapping zone, there is a
smooth transition from an avoided-crossing regime that gives rise
to a turbulent and complex $\lambda$ evolution to another regime
without much interaction between the eigenvalues that gives rise to a
smooth $\lambda$ evolution.

\subsection{Spacing distribution}

The eigenvalue diagrams only give us a qualitative understanding
of the fine energy structure of the spectrum. For a more
quantitative study, it is interesting to look at the spacing
distribution of the spectrum.

When $\lambda=0$, each semicircle distribution, corresponding to
a different system level, has a Wignerian level spacing
distribution
\begin{equation}
P^w(s) = \frac{\pi}{2} s e^{- \frac{\pi}{4} s^2}.
\end{equation}
But the total spectrum is made by the superposition, with a shift
$\Delta$, of two of such semicircle distributions. The shift
creates a Poissonian component to the total spacing distribution in
the overlapping zone ($\frac{\Delta}{2}-1$ to
$-\frac{\Delta}{2}+1$). A Poissonian distribution is given by
\begin{equation}
P^p(s) = e^{-s}.
\end{equation}
Therefore, we choose to fit the total spacing distribution by the
mixture
\begin{equation}
P^{fit}(s) = C_1 P^w(s) + (1-C_1) P^p(s). \label{fit}
\end{equation}
This choice of the form of the fit is empirical but reasonable because the
correlation coefficient of the fit is always close to one (between
$0.963$ and $0.982$).

\begin{figure}[h]
\centering 
\rotatebox{0}{\scalebox{0.4}{\includegraphics{fig13.eps}}} \\
\vspace*{1cm}
\rotatebox{0}{\scalebox{0.4}{\includegraphics{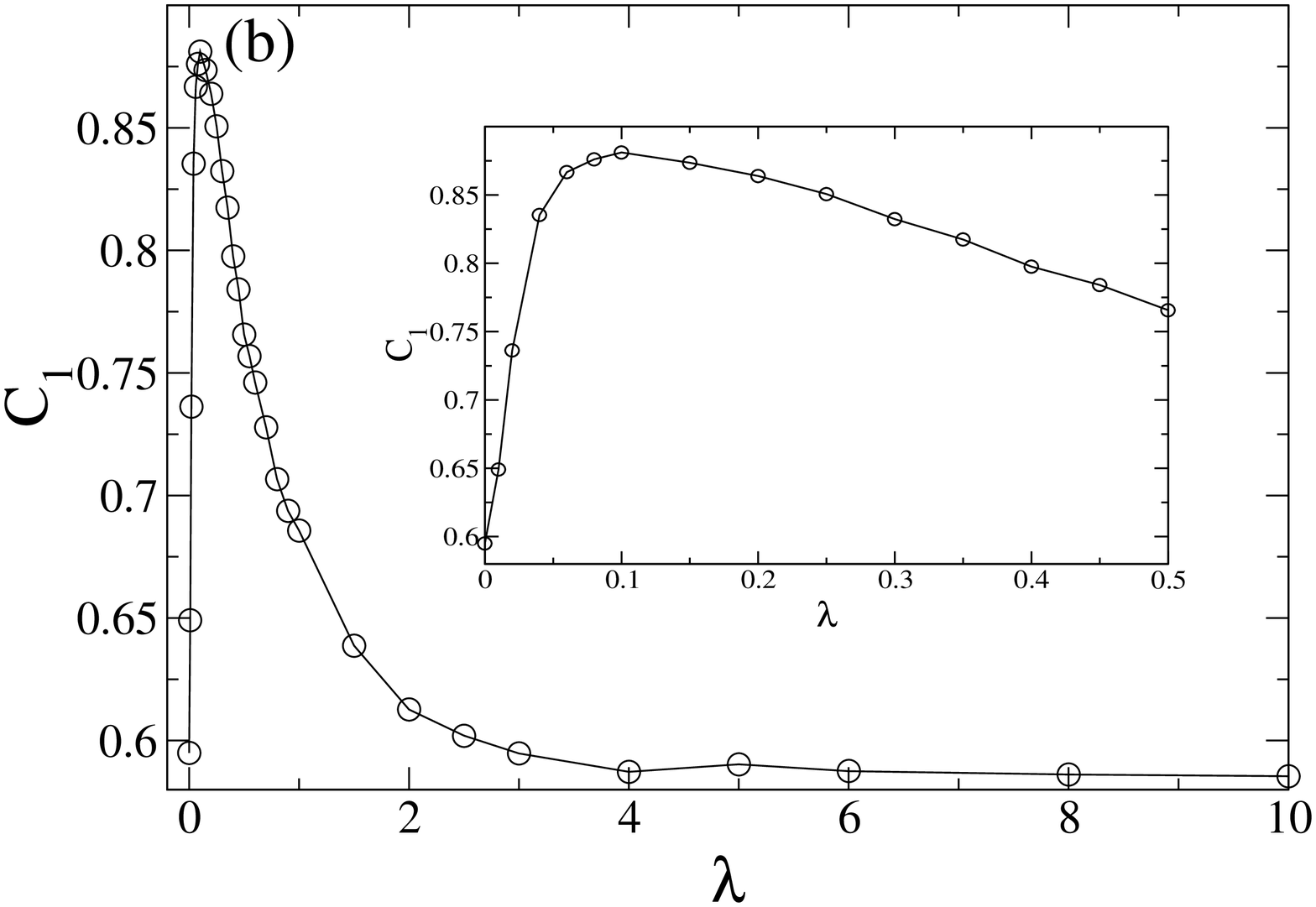}}}\\
\caption{(a) Spacing distribution for different
values of the coupling parameter $\lambda$. (b)
Fitted coefficient $C_1$ of Eq. (\ref{fit}). The closer is $C_1$ from
unity, the closer is the spacing distribution to the Wigner spacing
distribution. In the two figures, $\Delta=0.01$, $N=500$, and $\chi=50$.} 
\label{distribwigner}
\end{figure}

We computed the spacing distribution and made the
fit (\ref{fit}) in order to compute the mixing coefficient $C_1$ for
different values of $\lambda$ in the case where the overlapping
zone covers almost the whole spectrum ($\Delta \ll 1$). The results
are plotted in Fig. \ref{distribwigner}. We see that there is a
specific region of $\lambda$ values where the total spacing
distribution is close to a pure Wignerian one. This region
corresponds to the situation where $\lambda^2 N =O(1)$, i.e., when
the typical intensity of the interaction between the nonperturbed
levels is $O(\lambda^2)$ (since the first nonzero correction
in perturbation theory is of second order due to the nondiagonal
coupling in our model), becomes of the order of the mean level
spacing $O(\frac{1}{N})$ in the total system. In this
region, the level repulsion is maximal and effective among almost
all the states. 

In the limit $\lambda \to \infty$ the spectrum is
again the superposition of two semicircle distributions, one for the $E_{1
\eta}$'s and one for the $E_{-1 \eta}$'s according to Eq. (\ref{eqVPHoinv}).
As a consequence, one gets a Poissonian type of spacing distribution.

The other $\Delta$ cases
will have a Wignerian spacing distribution outside the overlapping
zone and a mixed one (like for the case $\Delta \ll 1$) inside.
This can be visually seen on the eigenvalue diagram that we
studied before.

\subsection{The shape of the eigenstates (SOE)}

We want now to have some information about the eigenstates of the
total system inside the overlapping zone of the two semicircle densities.
Following \cite{Cohen1}, we define the quantity
\begin{eqnarray}
\xi(\varepsilon,\varepsilon^0)=\sum_{\alpha,n} \vert \langle \alpha \vert n \rangle \vert^{2}
\delta(E_{\alpha}-\varepsilon) \delta(E_{n}-\varepsilon^{0})
\end{eqnarray}

If we fix $\varepsilon^{0}$ and study $\xi(\varepsilon,
\varepsilon^0)$ as a function of $\varepsilon$, we will call it
the local density of states (LDOS).

If we fix $\varepsilon$ and
study $\xi(\varepsilon,\varepsilon^0)$ as a function of
$\varepsilon^{0}$, we will call it the shape of the
eigenstates (SOE).

We here focus on the SOE. The SOE tells us how
close a perturbed eigenstate ($\lambda \neq 0$) at energy
$\varepsilon$ is from the nonperturbed eigenstate ($\lambda = 0$)
at energy $\varepsilon^0$. If the SOE is a very narrow function
centered around $\varepsilon = \varepsilon^0$, the concept of a
nonperturbed eigenstate is still useful. This regime corresponds
to very small coupling, for which the interaction intensity is
lower than the mean level spacing $0 < \lambda^2 N \lesssim 1$,
and will be called to the \textit{localized} regime. In the limit
$N \to \infty $ this regime disappears. If one increases the
coupling, the interaction intensity between the nonperturbed
states begins to be larger than the mean level spacing between the
states: $\lambda^2 N > 1$. The nonperturbed levels start to be
"mixed" by the interaction and the SOE starts then to have a
Lorentzian shape with a finite width $\Gamma$, centered around
$\varepsilon = \varepsilon^0$. This regime is called the
\textit{Lorentzian} regime. If one further increases the coupling
parameter, the SOE begins to spread over almost the whole spectrum.
This regime is called the \textit{delocalized} regime. 

In the banded random matrix model of \cite{Cohen1}, the regimes are classified
according to a different terminology and there is an additional
regime corresponding to a spreading that goes beyond the energy
range where the coupling acts (due to the finite coupling range of the
banded matrices). The motivation for our change of terminology will
become clear in the study of the dynamics.

\begin{figure}[h]
\centering
\rotatebox{0}{\scalebox{1}{\includegraphics{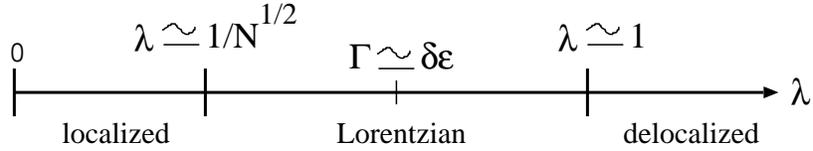}}} \\
\caption{Diagram of the different regimes as a function of the 
coupling parameter $\lambda$.}
\label{shemasoe}
\end{figure}

\begin{figure}[h]
\centering \rotatebox{0}{\scalebox{0.5}{\includegraphics{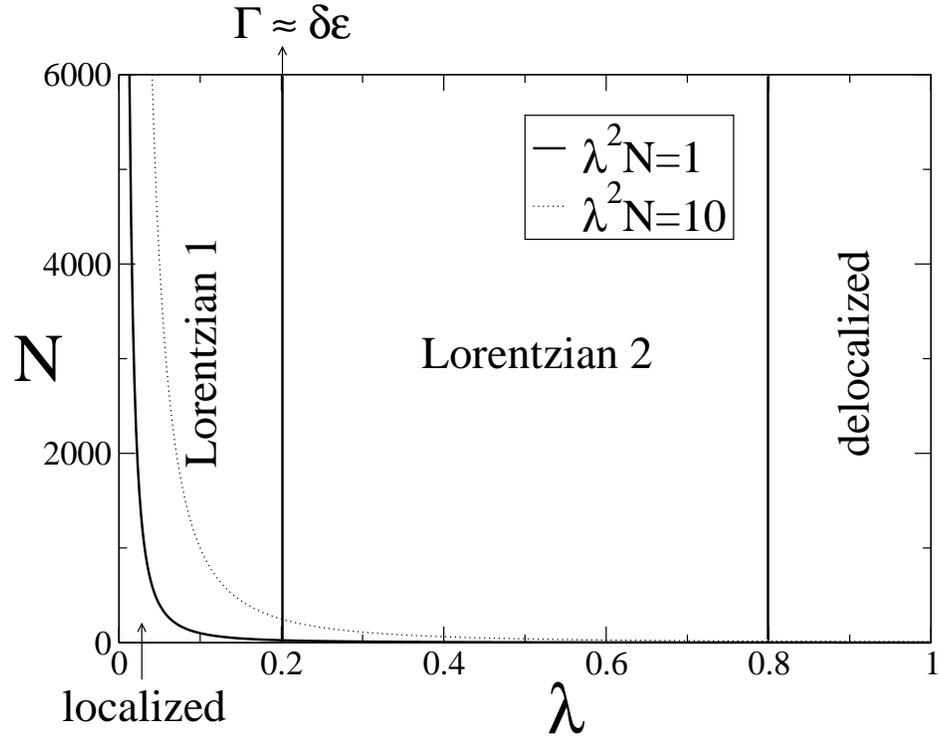}}}
\caption{Schematic representation of the different regimes in
the plane of the reduced parameter $\lambda$ and $N$. The Lorentzian $1$ 
regime corresponds to $\Gamma \lesssim \delta \epsilon$ and the Lorentzian 
$2$ regime to $\Gamma > \delta \epsilon$.}
\label{scheml2N}
\end{figure}

The Lorentzian regime can be separated into two parts. For small coupling,
the width of the Lorentzian $\Gamma$
is smaller or of the order of magnitude of the typical energy
scale of variation of the averaged smoothed density of state of
the environment $\delta \epsilon$: $n(\epsilon + \delta \epsilon)
\approx n(\epsilon)$. For larger coupling, the Lorentzian width extends 
on an energy scale larger than the typical energy
scale of variation of the density of states of the environment. 
Therefore, one has $\Gamma \lesssim \delta
\epsilon $ in the former case and $\Gamma > \delta \epsilon$  
in the latter case. The different regimes are represented 
in Figs. \ref{shemasoe} and \ref{scheml2N}.

\begin{figure}[h]
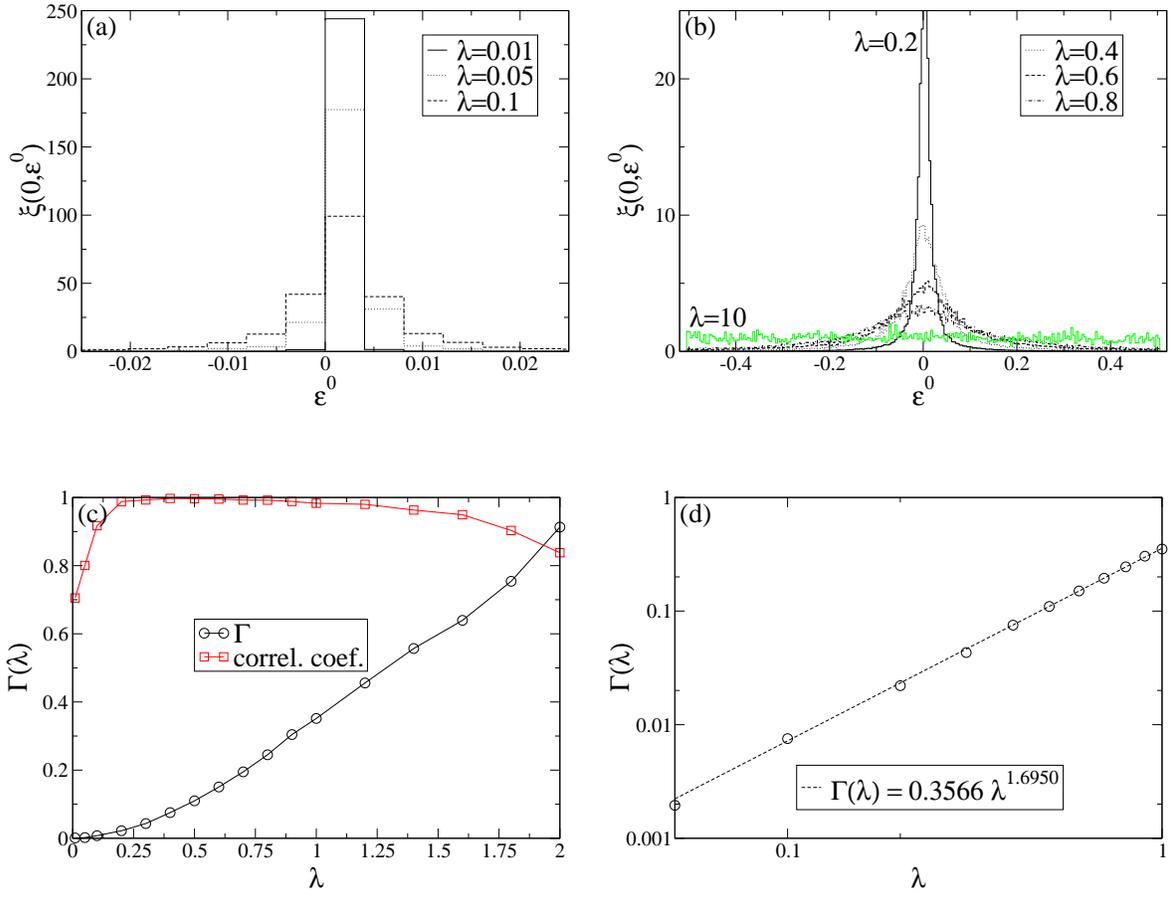

\centering
\begin{tabular}{c@{\hspace{0.5cm}}c}
\vspace*{1cm}
\rotatebox{0}{\scalebox{0.3}{\includegraphics{fig17.eps}}} &
\rotatebox{0}{\scalebox{0.3}{\includegraphics{fig18.eps}}} \\
\rotatebox{0}{\scalebox{0.3}{\includegraphics{fig19.eps}}} &
\rotatebox{0}{\scalebox{0.3}{\includegraphics{fig20.eps}}} \\
\end{tabular}
\caption{(a) SOE in the perturbative regime.
(b) SOE in the Lorentzian regime.
(c) The width and the correlation coefficient of the
fit of the SOE by a Lorentzian. (d) Power-law dependence of the width of
the Lorentzian SOE in the coupling
parameter. In all the figures: 
$\Delta=0.01$, $N=500$, $\chi=50$, and $\varepsilon=0$.} \label{SOEfigure}
\end{figure}

To represent the SOE of the spin-GORM model, we discretize the
energy axis in small cells of the order of the mean level spacing
$\frac{1}{N}$ and we average the SOE over $\chi$ realizations of
the random matrix ensemble. We see in 
Fig. \ref{SOEfigure}(a) the typical shape of the SOE going from
the perturbative regime ($\lambda=0.01,0.05$) to the beginning of
the Lorentzian one ($\lambda=0.1$). In Fig. \ref{SOEfigure}(b), we see the SOE
across the Lorentzian regime ($0.2 \leq \lambda \leq 0.8$). We also see 
the delocalized regime, when $\lambda \to \infty$ and the SOE gets 
completely flat ($\lambda=10$). Figure \ref{SOEfigure}(c) shows the
width of the Lorentzian from a fit made on the SOE curve. The
correlation coefficient of the fit helps us to determine the
region of the Lorentzian regime where the SOE is very well fitted by a
Lorentzian. It has been verified that the width of the Lorentzian
is independent of $N$ in the Lorentzian regime. Figure \ref{SOEfigure}(d) (log-log)
shows that the width of the Lorentzian has a power-law dependence
in the coupling parameter close to two in the Lorentzian regime.

\subsection{Asymptotic transition probability kernel (ATPK)}

An interesting quantity, which is close to the SOE, but which has
a nice physical interpretation, is the asymptotic transition
probability kernel (ATPK).

The transition probability kernel (TPK) gives the probability at
time $t$ to be in the level $\vert n \rangle$ if starting from $\hat{\rho}(0)$. 
It is defined as
\begin{eqnarray}
\Pi_{t}(n\vert\rho(0))=\langle n \vert  e^{-i \hat{H}_{\rm tot} t}
\hat{\rho}(0) e^{i \hat{H}_{\rm tot} t} \vert n \rangle
\end{eqnarray}
The ATPK is the time average of the TPK
\begin{eqnarray}
\Pi_{\infty}(n\vert\rho(0))&=& \lim_{T\to\infty} \frac{1}{T}
\int_{0}^{T} dt \Pi_{t}(n\vert\rho(0)) \nonumber \\ &=&
\sum_{\alpha} \vert \langle \alpha \vert n \rangle \vert^{2}
\langle \alpha \vert \hat{\rho}(0) \vert \alpha \rangle.
\end{eqnarray}
The distribution of the ATPK in energy is given by
\begin{eqnarray}
\Pi_{\infty}(\varepsilon^{0} \vert \rho(0)) = \sum_{n}
\Pi_{\infty}(n\vert\rho(0)) \delta(E_n-\varepsilon^{0}).
\end{eqnarray}
The ATPK has an intuitive physical interpretation. It represents
the probability after a very long time to end up in a nonperturbed
state $\vert n \rangle$, having started from the initial condition $\hat{\rho}(0)$. 
The ATPK is the convolution of the SOE
\begin{eqnarray}
\Pi_{\infty}(\varepsilon^{0} \vert \varepsilon^{0'}) = \sum_{n,m}
\sum_{\alpha} \vert \langle \alpha \vert n \rangle \vert^{2} \vert
\langle \alpha \vert m \rangle \vert^{2}
\delta(E_n-\varepsilon^{0}) \delta(E_m-\varepsilon^{0'}),
\end{eqnarray}
with $\rho(0) = \sum_m \vert m \rangle
\langle m \vert \delta(E_m-\varepsilon^{0'})$.

Because we are interested in a random matrix model and 
for the purpose of studying the dynamics, we will
perform averages of two different kinds. The first kind of average
is a microcanonical average over states belonging to the same
given energy shell of width $\delta \varepsilon$ for a given
realization of the random matrix ensemble used in our total
Hamiltonian. The second kind of average is an ensemble average
over the $\chi$ different realizations of the random matrices
ensemble. For the microcanonical average, 
a choice of the width of the energy shell
$\delta \varepsilon$ has to be
done in such a way that it is large enough to contains many levels 
(to get a good statistics) and
small enough to be smaller than or equal to the typical energy scale of
variation of the averaged smoothed density of states of the
environment $\delta \epsilon$. Therefore, the adequate choice of
the width of the energy shell corresponds to $\frac{1}{N} <
\delta \varepsilon < \delta \epsilon$.

\begin{figure}[h]
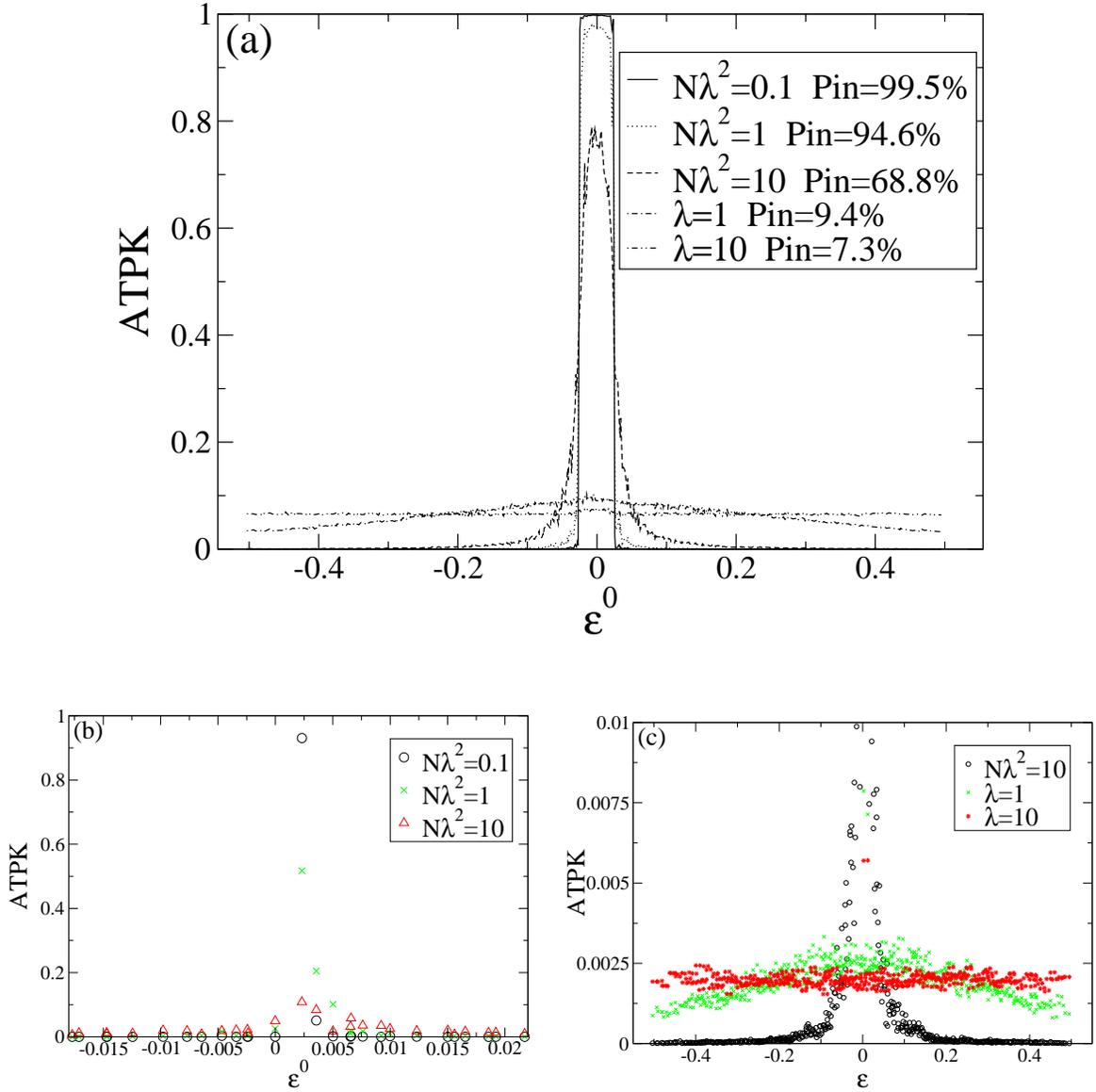

\centering
\rotatebox{0}{\scalebox{0.5}{\includegraphics{fig21.eps}}}\\
\centering \vspace*{1cm}
\begin{tabular}{c@{\hspace{0.5cm}}c}
\rotatebox{0}{\scalebox{0.3}{\includegraphics{fig22.eps}}} &
\rotatebox{0}{\scalebox{0.3}{\includegraphics{fig23.eps}}} \\
\end{tabular}
\caption{(a) ATPK from the localized
regime, through the Lorentzian regime, to the delocalized regime. The
ATPK is microcanonically averaged over the different initial
conditions corresponding to the levels inside the energy shell
centered at $\varepsilon^{0'}=0$ with width $\delta \varepsilon^{0'}=0.05$. 
"Pin" denotes the proportion of the ATPK that stays inside the initial energy 
shell after an infinite time. (b) and (c) ATPK for a single level as
an initial condition, without any average. All the figures are obtained
for very small system energy $\Delta=0.01$ with no random matrix ensemble 
average $\chi=1$ and for $N=500$.} \label{ATPKdessin}
\end{figure}

Different ATPK are depicted in Fig. \ref{ATPKdessin} where 
there is no random matrix ensemble average $\chi=1$. In Fig. \ref{ATPKdessin}(a), 
the ATPK is a microcanonical average inside the energy shell at energy
$\varepsilon^{0'}$ and of width $\delta \varepsilon^{0'}$. "Pin"
denotes the total probability of staying inside the energy
shell after a very long time. $N \lambda^2=1$ is on the
border between the localized and the Lorentzian regime
and $\lambda=1,10$ in the delocalized regime. We
see that, until $N \lambda^2=1$, the main probability stays in the
initial energy shell. But when the Lorentzian regime
starts, the probability spreads over energies larger than $\delta
\epsilon$ (here $\delta \epsilon \approx \delta
\varepsilon^{0'}$). In Figs. \ref{ATPKdessin}(b) and (c), no average has been 
done. The initial condition is a nonperturbed pure state (corresponding to
an energy close to zero) and one can see the individual
probability of being on another nonperturbed state for the
different regimes. We can see that in the localized regime the
probability of staying on the initial state is much more important
than the probability of leaving it. We also see that the Lorentzian regime
starts when the initial state loses its privileged position
containing the main probability and, therefore, when the neighboring
levels begin to have an important fraction of the total probability.

\section{Time evolution}
\label{time}

We now want to understand the time evolution of our model and more
specifically the population dynamics of the system.

The exact evolution of the total system is described by the
\textit{von Neumann equation}
\begin{equation}
\dot{\hat{\rho}}(t)=- i \lbrack \hat{H}_{\rm tot} , \hat{\rho}(t)
\rbrack , \label{von_Neumann}
\end{equation}
where $\hat{H}_{\rm tot}$ is given by Eq. (\ref{hamiltonien}). The
system dynamics is obtained from (\ref{von_Neumann}) using the
reduced density matrix $\hat{\rho}_S(t)=\textrm{Tr}_{B}
\hat{\rho}(t)$. The total system has a finite and constant
energy. At initial time, the environment has a given fixed energy
$\epsilon$ corresponding to a microcanonical distribution inside a
energy shell centered at $\epsilon$ and of width $\delta
\epsilon$. $\delta \varepsilon$ is chosen in such a way that it
is large enough to contains many levels (to get a good statistics)
and small enough to be smaller than or equal to the typical energy
scale of variation of the averaged smoothed density of state of
the environment $\delta \epsilon$. Therefore, the adequate choice
for the width of the energy shell corresponds to $\frac{1}{N} <
\delta \varepsilon < \delta \epsilon$. The dynamics is then
averaged over the $\chi$ realizations of the random matrix
ensemble.
In the following, we consider in detail the two
different extreme cases of weak and strong couplings.

\subsection{The weak coupling regime ($\lambda \ll 1$)}

We derived in Ref. \cite{Esposito} a perturbative equation for the
description of the evolution of a system (with a discrete spectrum)
weakly interacting with its environment (with a quasicontinious
spectrum). This equation has been shown to be equivalent to the
well-known \textit{Redfield equation} \cite{Redfield,GaspRed,Weiss} when the typical energy
scale of the system (typical energy spacing between the system
levels) can be considered small compared to the typical
environment energy scale $\delta \epsilon$ (typical energy scales
on which the smoothed density of states of the environment
varies). In the spin-GORM model, this condition means that
$n(\epsilon+\Delta) \approx n(\epsilon)$.

We will not perform in this paper the detailed derivation of this
equation and its application to the spin-GOE model. This has been
done in Ref. \cite{Esposito}. We will simply recall the main ideas and
results of this paper and apply them to the study of the
dynamics of our model.

The main idea is to suppose that the total density matrix can be
described at all time by a density matrix of the following form:
\begin{equation}
\hat{\rho}(t)=\frac{1}{n(\hat H_B)} \sum_{s,s'} \vert s \rangle \langle s' \vert P_{ss'}(\hat{H}_B;t),
\label{formegenrhopauli}
\end{equation}
where the environment density of states is
\begin{equation}
n(\epsilon)=\textrm{Tr}_{B} \delta(\epsilon-\hat{H}_B).
\end{equation}
Doing this, we neglect the contribution to the dynamics coming
from the coherence of the environment but we keep those of the
system. Therefore, the total density matrix is diagonal in the
environment degrees of freedom but not in the system ones. For
comparison, let us recall that in the derivation of the well known Pauli equation,
both types of coherences are neglected and the total density matrix
is completely diagonal \cite{Pauli,Zwanzig,Zubarev}.

The matrix of elements $P_{ss'}(\hat{H}_B;t)$ is Hermitian,
\begin{equation}
P_{s s'}(\hat{H}_B;t)=P_{s' s}^{\ast}(\hat{H}_B;t). \label{Phermi}
\end{equation}
The reduced density matrix of the system becomes
\begin{equation}
\hat{\rho}_S(t)=\textrm{Tr}_{B} \hat{\rho}(t)=
\int d\epsilon \textrm{Tr}_{B} \delta(\epsilon-\hat{H}_B) \hat{\rho}(t)=\sum_{s,s'}
\vert s \rangle \langle s' \vert \int d\epsilon P_{ss'}(\epsilon;t) .\label{rhospaulimic}
\end{equation}
We see that each element of the reduced density matrix of the system
depends on the environment energy. This is fundamental in order to
take into account the finite energy effects of the total system.

Using Eq. (\ref{formegenrhopauli}) and performing a perturbative
expansion up to the second order in $\lambda$ on Eq.
(\ref{von_Neumann}) (see Ref. \cite{Esposito}), one gets for
the population dynamics the equation
\begin{eqnarray}
\dot{P}_{ss}(\epsilon;t) = -2\lambda^2 \sum_{\bar{s},\bar{s}'}
\Big\lbrack && \langle s \vert
\hat{S} \vert \bar{s}' \rangle \langle \bar{s}' \vert \hat{S}
\vert \bar{s} \rangle P_{\bar{s}s}(\epsilon;t) \int d\epsilon'
F(\epsilon,\epsilon') n(\epsilon') \frac{\sin
(E_{\bar{s}}-E_{\bar{s}'}+\epsilon-\epsilon')\tau}{E_{\bar{s}}-E_{\bar{s}'}+
\epsilon-\epsilon'} \nonumber \\ &&-\langle s
\vert \hat{S} \vert \bar{s} \rangle \langle \bar{s}' \vert \hat{S}
\vert s \rangle n(\epsilon) \int d\epsilon' F(\epsilon,\epsilon')
P_{\bar{s}\bar{s}'}(\epsilon';t) \frac{\sin
(E_{s}-E_{\bar{s}'}+\epsilon-\epsilon')\tau}
{E_{s}-E_{\bar{s}'}+\epsilon-\epsilon'} \Big\rbrack \label{pauligenNM} ,
\end{eqnarray}
where $F(\epsilon,\epsilon')=``\vert \langle \epsilon \vert \hat{B} \vert \epsilon' \rangle
\vert^2"$, where the quotes denote a smoothening over a dense spectrum
of eigenvalues around $\epsilon$ and $\epsilon'$.
We see that the probability
$\dot{P}_{\bar{s}\bar{s}}(\epsilon;t)$, if initially concentrated
in a given energy shell, can spread in energy under the dynamics.
This is a typical non-Markovian effect happening on a short-time
scale. It is due to the presence in the equation of the energy
integral and of the $\frac{\sin(\xi \tau)}{\xi}$ function that has
a finite width in energy at short time. This equation is
non-Markovian in the sense that the coefficients of the
differential evolution equation are time-dependent and that this
time dependence can be neglected on long-time scales (performing
the Markovian approximation) when the environment has a faster
dynamics (i.e., typically a larger energy scale) than the system. 

The \textit{Markovian approximation} consists of taking the
infinite-time limit of the time dependent coefficients using the
property $\lim_{\tau \to \infty} \frac{\sin(\xi \tau)}{\xi}= \pi
\delta(\xi)$. This approximation is justified if the contribution
of the $\xi=0$ value (if it exists) has the main and almost unique
contribution to the energy integral. If one further neglects the
contributions of the coherence to the population evolution (this
is automatically satisfied for the spin-GORM model because of the
nondiagonal coupling), one gets a Pauli-type equation 
\cite{Pauli,Zwanzig,Zubarev}
\begin{eqnarray}
\dot{P}_{ss}(\epsilon;t) &=& -2\pi \lambda^2 \sum_{s'\ne s} \vert
\langle s \vert \hat{S} \vert s' \rangle \vert^2 
F(\epsilon,E_{s}-E_{s'}+\epsilon)
n(E_{s}-E_{s'}+\epsilon) P_{ss}(\epsilon;t) \nonumber \\ & &+2\pi
\lambda^2 \sum_{s'\ne s} \vert \langle s \vert \hat{S} \vert s'
\rangle \vert^2 F(\epsilon,E_{s}-E_{s'}+\epsilon) n(\epsilon)
P_{s's'}(E_{s}-E_{s'}+\epsilon;t) \label{origpaulipop}.
\end{eqnarray}
We see that the Markovian approximation strictly keeps the dynamics
of the total system inside an energy shell. Starting with the
probability located on a given energy shell, the dynamics will
preserve the probability inside this shell. But of course, the
probability of the different states inside the shell are varying.
The dynamics described by this equation can be seen as 
a random walk between nonperturbed states of the total system
belonging to the same energy shell with transition rates between
these states given by the Fermi golden rule.

We now apply our equation to the spin-GORM model in order to
study of the population evolution through $\hat{\sigma}_z$
(the difference between the probability of being in the upper state of the
system minus the probability of being in the lower one). Doing
this, we suppose that the environment is quasi-continuous $N \to \infty$
and that the random matrix ensemble average has been performed
$\chi \to \infty$. For the non-Markovian equation (\ref{pauligenNM}), one gets
\begin{eqnarray}
\langle \hat{\sigma}_z \rangle^{NM}(t) = \int d\epsilon' \left[
P_{++}(\epsilon';t)-P_{--}(\epsilon'+\Delta;t) \right] ,
\label{zdefpauligen1}
\end{eqnarray}
where
\begin{eqnarray}
\dot{P}_{\pm \pm}(\epsilon;t) &=& \frac{\lambda^2}{\pi}
\int_{-\frac{1}{2}}^{+\frac{1}{2}} d\epsilon'
\frac{\sin(\pm \Delta+\epsilon-\epsilon')t}{(\pm \Delta+\epsilon-\epsilon')}
\left[ P_{\mp \mp}(\epsilon';t) \sqrt{\frac{1}{4}-{\epsilon}^2} -
P_{\pm \pm}(\epsilon;t) \sqrt{\frac{1}{4}-{\epsilon'}^2} \right].
\label{paulispingoeNM11}
\end{eqnarray}

In the Markovian limit, Eq. (\ref{zdefpauligen1}) becomes
\begin{eqnarray}
\langle \hat{\sigma}_z \rangle^{M}(t) =
P_{++}(\epsilon;t)-P_{--}(\epsilon+\Delta;t),
\label{zdefpauligen2}
\end{eqnarray}
which obeys a Pauli-type equation so that
we get
\begin{eqnarray}
\langle \hat{\sigma}_z \rangle^{M}(t)= \langle \hat{\sigma}_z
\rangle^{M}_{\infty}+ \left[ \langle \hat{\sigma}_z
\rangle^{M}(0) - \langle \hat{\sigma}_z \rangle^{M}_{\infty}
\right] e^{- \gamma t}, \label{paulispingoeMZt}
\end{eqnarray}
where the equilibrium value of the populations is given by
\begin{equation}
\langle \hat{\sigma}_z \rangle^{M}_{\infty}
=\frac{\sqrt{\frac{1}{4}-(\epsilon)^2}-
\sqrt{\frac{1}{4}-(\epsilon+\Delta)^2}}
{\sqrt{\frac{1}{4}-(\epsilon)^2} +
\sqrt{\frac{1}{4}-(\epsilon+\Delta)^2}} 
\label{paulispingoeMZinfini}
\end{equation}
and the relaxation rate by
\begin{equation}
\gamma= \lambda^2 \left( \sqrt{\frac{1}{4}-(\epsilon)^2}+
\sqrt{\frac{1}{4}-(\epsilon+\Delta)^2}
\right) ,\label{paulispingoeMrate}
\end{equation}
where we adopted the convection that $\sqrt{x}=0$ if $x<0$.

In the case of the spin-GORM model, the transition probability
between states belonging to the same total energy shell only depends
on the smoothed density of states. This
is due to the fact that the environment coupling elements between
the environment nonperturbed states are randomly distributed
because the environment coupling operator is a random matrix. This
has the consequence that the equilibrium value of the populations
(\ref{paulispingoeMZinfini}) corresponds to a microcanonical
distribution probability of the states belonging to the total
energy shell independently of the initial distribution of these
states inside this energy shell.

We notice that, in the general case, the Markovian assumption made on
our equation does not directly give a Pauli-type equation which is
an equation for the populations only. To get a Pauli equation, the further
approximation, which consists of neglecting the contribution of the
coherences to the population dynamics, has to be done. For the
particular case of the spin-GORM model, this further approximation
is not necessary because it is automatically satisfied.

\subsection{The strong coupling regime ($\lambda \gg 1$)}

We are now interested in describing the dynamical regime where
the coupling parameter $\lambda$ is very large and, therefore,
dominant in front of $1$ and
$\Delta$. We will again suppose that the environment is
continuous $N \to \infty$ and that the random matrix ensemble
average has been performed $\chi \to \infty$.

The population dynamics of the system is given by
\begin{eqnarray}
\langle \hat{\sigma}_z \rangle(t)&=& \textrm{Tr} \ \ \hat{\rho}(t)
\hat{\sigma}_z \\ &=&\textrm{Tr} \ \ e^{i
(\frac{\Delta}{2}\hat{\sigma}_z+\hat{H}_{B}+\lambda\hat{\sigma}_x\hat{B})
t} \hat{\rho}(0) e^{-i
(\frac{\Delta}{2}\hat{\sigma}_z+\hat{H}_{B}+\lambda\hat{\sigma}_x\hat{B})
t} \hat{\sigma}_z. \nonumber
\end{eqnarray}
Using the following unitary transformation acting on the spin degrees of freedom
\begin{equation}
\hat{U}=
\left(\begin{array}{cc}
\frac{1}{\sqrt{2}} & \frac{1}{\sqrt{2}} \\
\frac{1}{\sqrt{2}} & -\frac{1}{\sqrt{2}}
\end{array} \right) \label{rhoSexplicite}
\end{equation}
we get
\begin{eqnarray}
\langle \hat{\sigma}_z \rangle(t)&=& \textrm{Tr} \ \ e^{i
(\frac{\Delta}{2}\hat{\sigma}_x+\hat{H}_{B}+\lambda\hat{\sigma}_z\hat{B})
t} \hat{U}^{\dag} \hat{\rho}_{S}(0) \hat{U} 
\hat{\rho}_{B}(0) e^{-i
(\frac{\Delta}{2}\hat{\sigma}_x+\hat{H}_{B}+\lambda \hat{\sigma}_z
\hat{B}) t}  \hat{\sigma}_x \nonumber \\ &=& \textrm{Tr} \ \ e^{i
\lambda (\frac{1}{\lambda}\frac{\Delta}{2}\hat{\sigma}_x+
\frac{1}{\lambda} \hat{H}_{B} + \hat{\sigma}_z \hat{B}) t}
\hat{U}^{\dag} \hat{\rho}_{S}(0) \hat{U} 
\hat{\rho}_{B}(0) e^{-i \lambda (\frac{1}{\lambda}
\frac{\Delta}{2} \hat{\sigma}_x+ \frac{1}{\lambda}\hat{H}_{B}
+\hat{\sigma}_z \hat{B}) t} \hat{\sigma}_x .
\end{eqnarray}
Using the following perturbative expansion of the evolution
operator to order zero in $\frac{1}{\lambda}$:
\begin{eqnarray}
e^{-i \lambda (\frac{1}{\lambda} \frac{\Delta}{2}\hat{\sigma}_x +
\frac{1}{\lambda}\hat{H}_{B}+ \hat{\sigma}_z \hat{B}) t}
&\stackrel{\frac{1}{\lambda} \to 0}{=}& e^{-i \lambda t
\hat{\sigma}_z \hat{B}} + O\left(\frac{1}{\lambda}\right),
\end{eqnarray}
we get
\begin{eqnarray}
\langle \hat{\sigma}_z \rangle(t)&\stackrel{\frac{1}{\lambda} \to
0}{=}& \textrm{Tr} \ \ e^{i \lambda t \hat{\sigma}_z \hat{B}}
\hat{U}^{\dag} \hat{\rho}_{S}(0) \hat{U} 
\hat{\rho}_{B}(0) e^{-i \lambda t \hat{\sigma}_z \hat{B}}
+O\left(\frac{1}{\lambda}\right).
\end{eqnarray}

Using the following notation
\begin{eqnarray}
\hat{\sigma}_z \hat{B} \vert \kappa \eta \rangle = \hat{\sigma}_z
\vert \kappa \rangle \hat{B} \vert \eta \rangle = \kappa
E_{\eta} \vert \kappa \eta \rangle
\end{eqnarray}
and
\begin{eqnarray}
\hat{H}_B  \vert b \rangle =  E_b^{B} \vert b \rangle,
\end{eqnarray}
we find that
\begin{eqnarray}
\langle \hat{\sigma}_z \rangle(t) &\stackrel{\frac{1}{\lambda} \to
0}{=}& \sum_{\kappa,\eta} e^{i \lambda t \kappa E_{\eta} } \langle
\eta \vert \hat{\rho}_B(0) \vert \eta \rangle \langle \kappa \vert
\hat{U}^{\dag} \hat{\rho}_{S}(0) \hat{U} \vert -\kappa \rangle
e^{i \lambda t \kappa E_{\eta} } + O\left(\frac{1}{\lambda}\right).
\label{szprov1}
\end{eqnarray}
Because $\hat{\rho}_B$ is diagonal in the basis that diagonalizes
$\hat{H}_{B}$, we have
\begin{eqnarray}
\langle \eta \vert \hat{\rho}_B(0) \vert \eta \rangle = \sum_{b}
\vert \langle \eta \vert b \rangle \vert^2 \langle b \vert
\hat{\rho}_B \vert b \rangle.
\end{eqnarray}
If we perform an ensemble average over different realizations of
$\hat{H}_{B}$ and use the random matrix eigenvectors statistics
\cite{Mehta,Porter,Brody}, we find
\begin{eqnarray}
\overline{\langle \eta \vert \hat{\rho}_B(0) \vert \eta
\rangle}^{(\hat{H}_{B})} &=& \sum_{b} \overline{\vert \langle \eta
\vert b \rangle \vert^2}^{(\hat{H}_{B})} \overline{\langle b \vert
\hat{\rho}_B \vert b \rangle}^{(\hat{H}_{B})} \nonumber \\
&=&\sum_{b} \frac{2}{N} \overline{\langle b \vert \hat{\rho}_B
\vert b \rangle}^{(\hat{H}_{B})} = \frac{2}{N}.
\end{eqnarray}
Equation (\ref{szprov1}) therefore becomes
\begin{eqnarray}
\overline{\langle \hat{\sigma}_z \rangle(t)}^{(\hat{H}_{B})}
&\stackrel{\frac{1}{\lambda} \to 0}{=}& \frac{2}{N}
\sum_{\kappa,\eta} e^{2i \lambda t \kappa E_{\eta} } \langle
\kappa \vert \hat{U}^{\dag} \hat{\rho}_{S}(0) \hat{U} \vert
-\kappa \rangle \\ &=& \frac{2}{N} \left[\langle 1 \vert \hat{U}^{\dag}
\hat{\rho}_{S}(0) \hat{U} \vert -1 \rangle \sum_{\eta} e^{2i
\lambda t E_{\eta} } + \langle -1 \vert \hat{U}^{\dag}
\hat{\rho}_{S}(0) \hat{U} \vert 1 \rangle) \sum_{\eta} e^{-2 i
\lambda t E_{\eta} }\right] + O\left(\frac{1}{\lambda}\right). \nonumber
\end{eqnarray}
Performing now the following ensemble average over different
realizations of $\hat{B}$:
\begin{eqnarray}
\overline{\sum_{\eta} e^{2i \lambda t E_{\eta}}}^{(\hat{B})} &=&
\int_{-\frac{1}{2}}^{+\frac{1}{2}} d\epsilon \ \ n(\epsilon) e^{2i
\lambda t \epsilon}  \nonumber \\ &=& \frac{4N}{\pi}
\int_{-\frac{1}{2}}^{+\frac{1}{2}} d\epsilon \ \
\sqrt{\frac{1}{4}-\epsilon^2} e^{2i \lambda t \epsilon} \nonumber
\\ &=& N \frac{J_{1}(\lambda t)}{\lambda t},
\end{eqnarray}
we finally get
\begin{eqnarray}
\overline{\langle \hat{\sigma}_z
\rangle(t)}^{(\hat{H}_{B},\hat{B})} &\stackrel{\frac{1}{\lambda}
\to 0}{=}& 2\frac{J_{1}(\lambda t)}{\lambda t} \left[\langle 1 \vert
\hat{U}^{\dag} \hat{\rho}_{S}(0) \hat{U} \vert -1 \rangle+ \langle
-1 \vert \hat{U}^{\dag} \hat{\rho}_{S}(0) \hat{U} \vert 1
\rangle\right]+ O\left(\frac{1}{\lambda^2}\right). \nonumber \\
\end{eqnarray}
It is easy to show that the term of order $\frac{1}{\lambda}$ is zero.
This explains the fact that $O\left(\frac{1}{\lambda}\right)$ has
been replaced by $O\left(\frac{1}{\lambda^2}\right)$.

Choosing as an initial condition
\begin{equation}
\hat{\rho}_{S}(0)=
\left(\begin{array}{cc}
1 & 0 \\
0 & 0
\end{array} \right) , \label{rhoSzeromat}
\end{equation}
we get
\begin{eqnarray}
\overline{\langle \hat{\sigma}_z
\rangle(t)}^{(\hat{H}_{B},\hat{B})}&\stackrel{\frac{1}{\lambda}
\to 0}{=}& 2\frac{J_{1}(\lambda t)}{\lambda t} +
O\left(\frac{1}{\lambda^2}\right) . \label{Besstlsurtl}
\end{eqnarray}
We have found a well-defined behavior of the
system dynamics when the coupling parameter $\lambda$ is so large
that the coupling term can be considered to contribute alone to the
dynamics.

\subsection{Numerical results}

We will now numerically study the validity of the approximated
equation that we just derived in order to understand the dynamical
evolution of our model.

In our numerical simulations, the initial condition of the system
is always the upper state (\ref{rhoSzeromat}): $\langle
\hat{\sigma}_z \rangle(0)=1$ . The different parameter domains
represented in Fig. \ref{regimeschema} will play a fundamental
role in our discussion of the dynamics.

\begin{figure}[p]
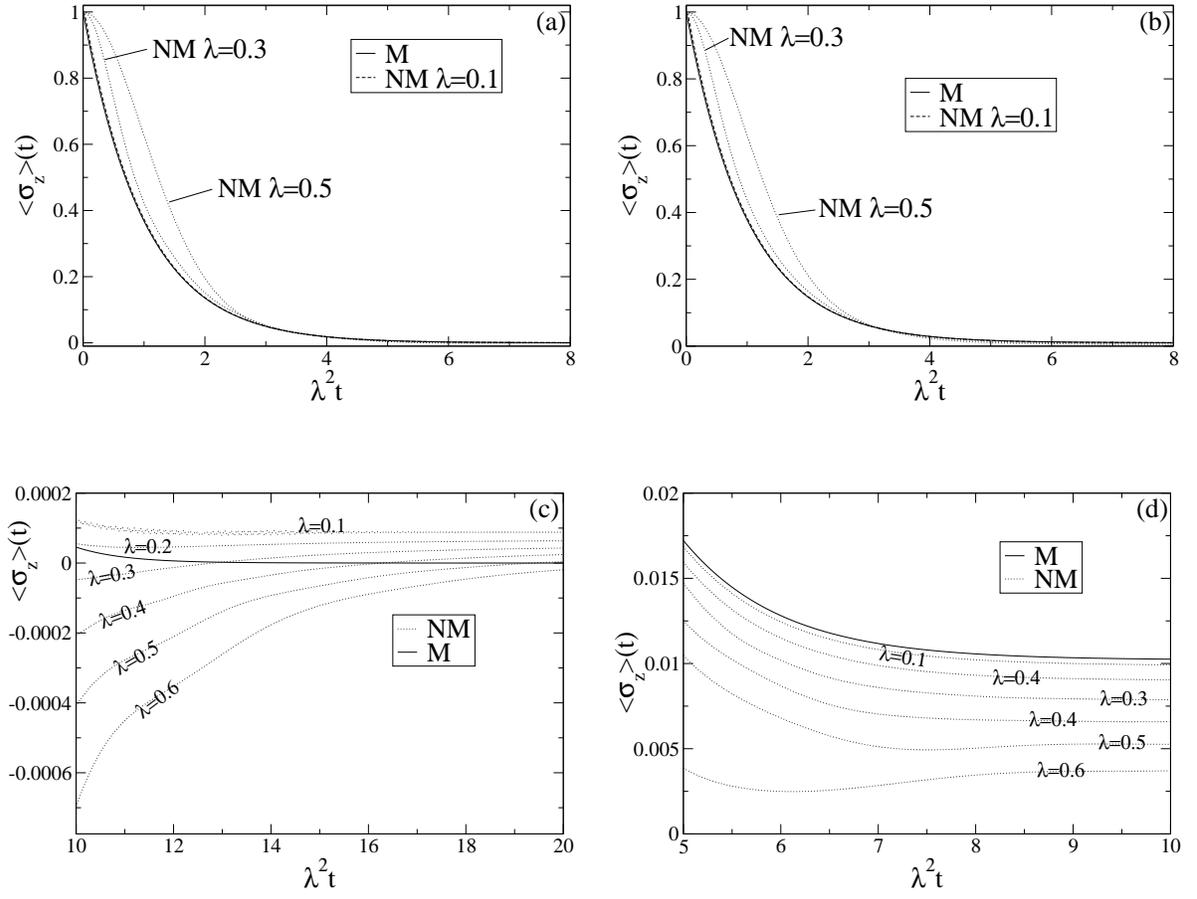

\centering
\begin{tabular}{c@{\hspace{0.5cm}}c}
\vspace*{1cm}
\rotatebox{0}{\scalebox{0.3}{\includegraphics{fig24.eps}}}
&
\rotatebox{0}{\scalebox{0.3}{\includegraphics{fig25.eps}}}
\\
\rotatebox{0}{\scalebox{0.3}{\includegraphics{fig26.eps}}}
&
\rotatebox{0}{\scalebox{0.3}{\includegraphics{fig27.eps}}}
\end{tabular}
\caption{Time evolution of the $z$ component of the spin: comparison between 
the Markovian (M) and non-Markovian (NM) versions of the perturbative equation 
for different values of the coupling parameter. In all the cases $\Delta=0.01$ 
and $\epsilon=0$.}
\label{PauliMvsNM}
\end{figure}

We begin by the comparison between the results of the non-Markovian 
version (\ref{zdefpauligen1}) and (\ref{paulispingoeNM11}) of our perturbative 
equation and the results of the Markovian version 
(\ref{paulispingoeMZt})-(\ref{paulispingoeMrate}) (of Pauli type) in order 
to understand better 
the consequences of the Markovian approximation.
Figures \ref{PauliMvsNM}(a) and \ref{PauliMvsNM}(b) show the time evolution
of $\langle\hat\sigma_z\rangle$ for both equations at different
values of the coupling parameter. The time axis has been scaled by
the coupling parameter ($\lambda^2 t$). This
scaling is characteristic of the Lorentzian SOE regime.
The values of $\lambda$ have of course to be reasonably small in
order to remain consistent with the fact that these equations are
obtained perturbatively. The time scale that we are observing is
the global one: from the initial condition to the equilibrium. The
characteristic energy of the system ($\Delta$) is different in
Figs. \ref{PauliMvsNM}(a) and \ref{PauliMvsNM}(b). But we are in both cases in domain A
of the reduced parameter phase space (see
Fig. \ref{regimeschema}). For such small values of $\Delta$, we see almost no
difference between Figs. \ref{PauliMvsNM}(a) and \ref{PauliMvsNM}(b). 
The characteristic time scale of the
environment is of order $2 \pi$, and is therefore in both cases
much shorter than the system one. We are in situations where the
Markovian approximation makes sense on times longer than $2 \pi$.
Using the $\lambda^2 t$ scaling, the Markovian equation is
independent of $\lambda$. This is not the case for the
non-Markovian equation. We see that the stronger $\lambda$ is, the
larger the deviation is between both equations. This is the
consequence of the fact that the $\lambda^2 t$ scaling that we use
amplifies the non-Markovian short time behavior (that occurs on
time of order of the characteristic time scale of the environment
$2 \pi$) when $\lambda$ increases.
Figs. \ref{PauliMvsNM}(c) and \ref{PauliMvsNM}(d) show us the long
time behavior of the Markovian and non-Markovian versions
of our equation. We see that the equilibrium value of $\langle
\hat{\sigma}_z \rangle$ depends on $\lambda$ for the non-Markovian
equation. This is not the case for the Markovian equation. The
differences between the equilibrium values are small. But,
comparing Fig. \ref{PauliMvsNM}(c) with \ref{PauliMvsNM}(d), one 
notices that when $\Delta$ is
larger, the differences are larger. This is a consequence of the
error made on short-time dynamics using the Markovian
approximation. Because this error is more important when $\Delta$
is large, the consequence on the long time dynamics is more
important, even if globally small.

\begin{figure}[p]
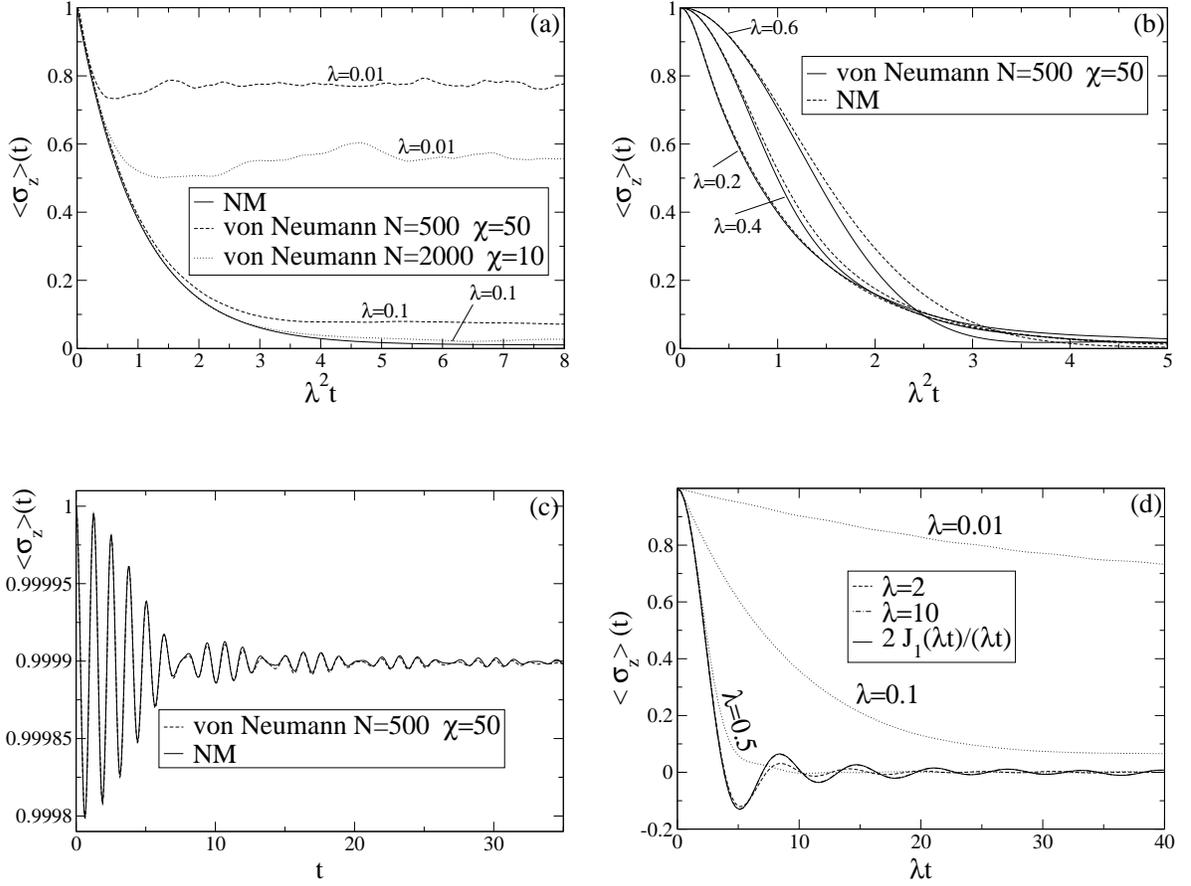

\centering
\begin{tabular}{c@{\hspace{0.5cm}}c}
\vspace*{1cm}
\rotatebox{0}{\scalebox{0.3}{\includegraphics{fig28.eps}}}
&
\rotatebox{0}{\scalebox{0.3}{\includegraphics{fig29.eps}}}
\\ 
\rotatebox{0}{\scalebox{0.3}{\includegraphics{fig30.eps}}}
&
\rotatebox{0}{\scalebox{0.3}{\includegraphics{fig31.eps}}} \\
\end{tabular}
\caption{Time evolution of the $z$ component of the spin. 
(a), (b), and (c) Comparison between the 
exact von Neumann equation and the non-Markovian (NM) 
version of the perturbative equation for different values of the 
coupling parameter. (a) and (b) correspond to $\Delta=0.1$ and 
(c) to $\Delta=0.5$ and $\lambda=0.1$.
(d) Comparison  between the exact von Neumann equation 
and the Bessel strong coupling result given by Eq. (\ref{Besstlsurtl}) 
for $\Delta=0.01$, $N=500$, and $\chi=50$. In all the figures $\chi=50$.} 
\label{differents regimes}
\end{figure}

We now compare the non-Markovian version 
(\ref{zdefpauligen1})-(\ref{paulispingoeNM11}) of our perturbative
equation with the exact von Neumann equation (\ref{von_Neumann}), staying 
in the parameter domain A. Figures \ref{differents regimes}(a) and 
\ref{differents regimes}(b) show the time evolution described by both equations at
different $\lambda$ values using the $\lambda^2 t$ time scaling. 
In Fig. \ref{differents regimes}(a), the Markovian and
the non-Markovian versions of our equation are so close that we
only plotted the second one. We see that the
non-Markovian equation is valid not only for values of
$\lambda$ below an upper bound, but also above a lower bound. 
When $\lambda$ is too small,
the non-Markovian version of our perturbative equation does not
fit with the exact result. The exact initial dynamics is well
reproduced by our equation, but the relaxation process
to the equilibrium value is not reproduced. This is due to the discrete
nature of the spectrum and therefore depends on the number
$N$ of states and disappears in the limit $N \to \infty$. It corresponds to
the border between the localized and Lorentzian regimes
of the SOE. This is one of the main results of this paper. This
phenomena is of course again related to the effect of the
perturbation between the levels in the total spectrum that we
already observed in the spacing distribution, in the SOE, and in
the ATPK. $\lambda$ has to be large enough ($\lambda^2 >
\frac{1}{N}$) to "mix" the nonperturbed levels in order
for all the states inside the total unperturbed
energy shell to be mixed together.  Remember
that the equilibrium value of our Markovian perturbative equation
is given by a microcanonical distribution in the total unperturbed 
energy shell (see Eq.(\ref{paulispingoeMZinfini})). One also sees 
in Fig. \ref{differents regimes}(b) that
our perturbative equation loses again its validity above a certain
value of $\lambda$. It happens at value of $\lambda$ that cannot
be considered as perturbative any more. It also corresponds to a
value of the coupling parameter corresponding in the SOE to the
transition in the Lorentzian regime when the width 
of the Lorentzian starts to be larger than
the typical variation energy scale of the environment density of
states.

In Fig. \ref{differents regimes}(c), we deal with the
parameter domain B (see Fig. \ref{regimeschema}). We again
compare the non-Markovian version of our perturbative equation
with the exact dynamics given by the von Neumann equation. In this
case, the system dynamics (of period $0.4 \pi$) is faster than the
environment one (of period $2 \pi$). We are in a highly non-Markovian
situation. The Markovian version of our equation (that describes no
evolution in this case) completely misses the observed behavior of
damped oscillations. But the non-Markovian equation reproduces
this behavior with a very high accuracy. The frequency of
the oscillations corresponds to the system dynamics ones and the
damping of these oscillations occur on a time scale corresponding
to the environment characteristic time scale.  

Finally, in Fig. \ref{differents regimes}(d), we are in
the high coupling parameter domain C (see
Fig. \ref{regimeschema}). We see that when $\lambda$ becomes large 
enough to make
the coupling term dominant in the total Hamiltonian, the dynamics
obey the Bessel behavior derived in Eq.(\ref{Besstlsurtl}). It is
important to notice that this behavior scales in time according to
$\lambda t$.

A summary of the validity of the different approximated equations
(Markovian (\ref{paulispingoeMZt})-(\ref{paulispingoeMrate}) and 
non-Markovian (\ref{zdefpauligen1}) and (\ref{paulispingoeNM11}) versions 
of our perturbative equation and strong coupling Bessel equation 
(\ref{Besstlsurtl})) and of the different scalings, depending on the 
regime that one considers, is represented in Fig. \ref{valideq}.

\subsection{Average versus individual realizations}

An interesting point is the comparison, for the system dynamics,
between the averaged (random matrix
ensemble averaged or microcanonically averaged) dynamics and the
dynamics of an individual realization within the statistical ensembles. 
This latter corresponds to a dynamics generated by an initial
condition that corresponds to a pure state and without any
random-matrix average ($\chi=1$).

\begin{figure}[h]
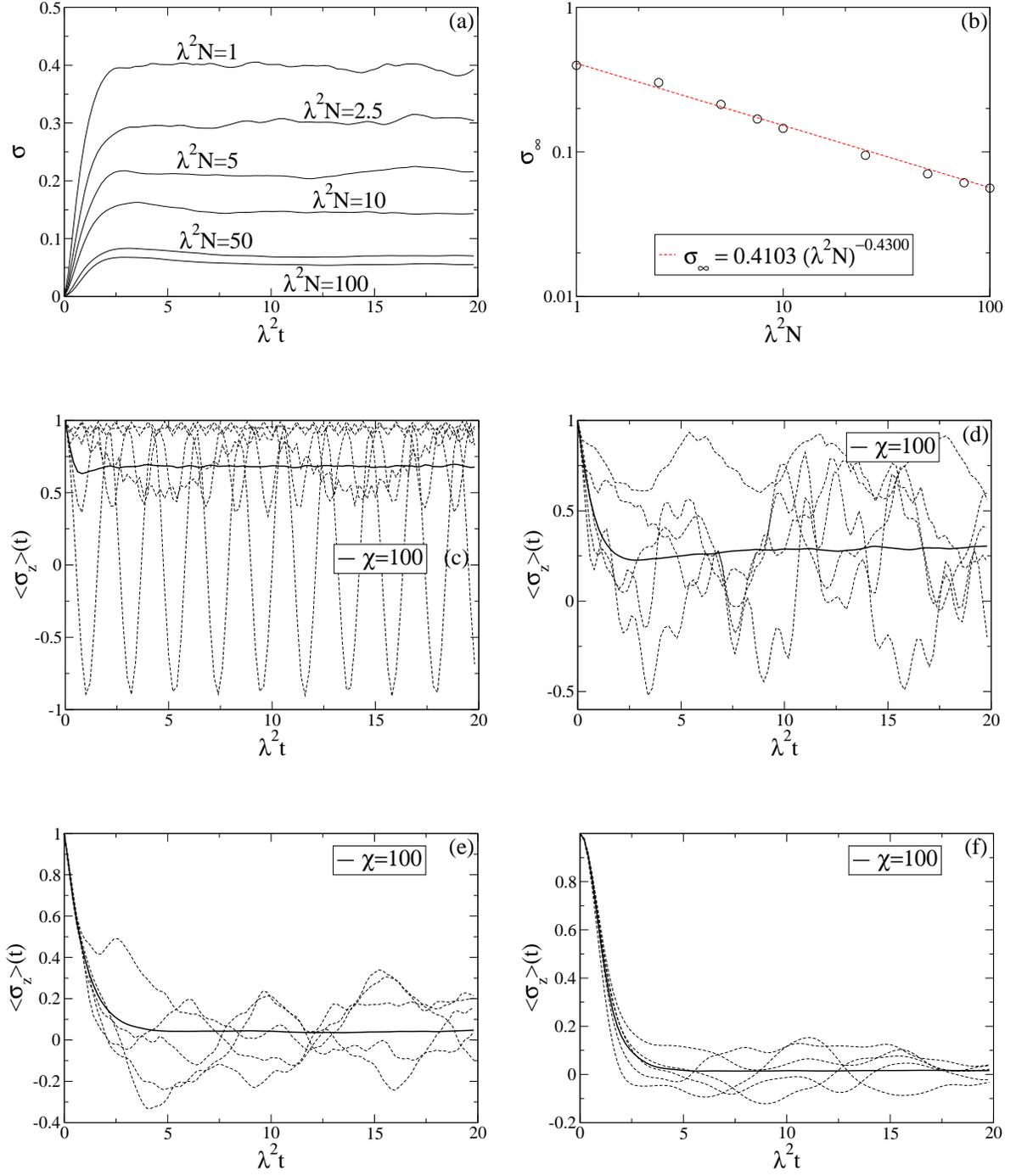

\centering
\begin{tabular}{c@{\hspace{0.5cm}}c}
\vspace*{1cm}
\rotatebox{0}{\scalebox{0.3}{\includegraphics{fig33.eps}}}
&
\rotatebox{0}{\scalebox{0.3}{\includegraphics{fig34.eps}}}
\\ \vspace*{1cm}
\rotatebox{0}{\scalebox{0.3}{\includegraphics{fig35.eps}}}
&
\rotatebox{0}{\scalebox{0.3}{\includegraphics{fig36.eps}}}
\\ 
\rotatebox{0}{\scalebox{0.3}{\includegraphics{fig37.eps}}}
&
\rotatebox{0}{\scalebox{0.3}{\includegraphics{fig38.eps}}}
\\
\end{tabular}
\caption{In all the figures $\Delta=0.1$, $N=500$, $\epsilon=0$, 
and $\delta \epsilon=0.05$. (a) Variance between individual
trajectories and the averaged one ($\chi=100$) as a function of 
time for different values of $\lambda^2 N$. (b) Power-law
dependence between the equilibrium value of this variance and
$\lambda^2 N$. (c)-(f) Individual
trajectories of the ensemble (dashed lines) and the ensemble
averaged trajectory (solid line). In (c) $\lambda^2 N=0.1$, in (d) 
$\lambda^2 N=1$, in (e) $\lambda^2 N=10$, and in (f) 
$\lambda^2 N=100$.} \label{Zfluct}
\end{figure}

We see in Fig. \ref{Zfluct}(c)-(f)
the dynamics of a system with small energy spacing $\Delta=0.1$
for different values of $\lambda^2 N$. The solid line represents the
random-matrix and microcanonically averaged
dynamics. The dashed lines depict some of the individual
members of the random-matrix ensemble corresponding to an initial
pure state inside the total unperturbed energy shell. We see that the larger
$\lambda^2 N$ is, the closer the individual trajectories are 
from the averaged trajectories. We therefore have, when $\lambda^2 N$ is
large enough, that the individual realizations are self-averaging
in the random-matrix and microcanonical ensembles. In
order to quantify this behavior, we plotted in Fig. \ref{Zfluct}(a) the variance between
the individual trajectories and the averaged trajectories as a function of
time for different values of $\lambda^2 N$. We observe that 
this variance decreases as $\lambda^2N\to\infty$. 
In Fig. \ref{Zfluct}(b), we show that the asymptotic value of the variance 
decreases with a power-law dependence with respect to $\lambda^2 N$.

We again see the relation with the SOE regimes. In the localized
regime (Fig. \ref{Zfluct}(c)), the dynamics is governed by very different
individual trajectories oscillating with a very few frequencies
that differ from one individual trajectory to another.
This is a consequence of the fact that the perturbed levels are
still close to the nonperturbed ones and are only slightly 
affected by neighboring nonperturbed levels. In the
Lorentzian regime (Figs. \ref{Zfluct}(d) and (e)), each individual trajectory
follows roughly the averaged trajectory and contains a very large
number of different frequencies. This shows that the interaction
"mixes" many of the nonperturbed levels, deleting the discrete
structure of the spectrum.

Therefore, we can say that our master equation (\ref{pauligenNM}) or
(\ref{origpaulipop}) holds with a given accuracy for a majority of individual 
trajectories if $\lambda$ is small enough satisfying $\lambda \geq C N^{-\nu}$ 
with $\nu <\frac{1}{2}$  and a constant $C>0$, in the limit $N\to\infty$.

\section{The very long time behavior}
\label{equilibrium}

We here focus on the very long time behavior of our model,
in other words on its equilibrium properties.

\subsection{Equilibrium values of the system observables} \label{valequiobssys}

Let us consider the spin observable $\hat{\sigma}_z$ of the
two-level system. This observable evolves in time according to
\begin{eqnarray}
\langle \hat{\sigma}_z \rangle(t)&=&\textrm{Tr} \ \  \hat{\rho}(0)
e^{i \hat{H}_{\rm tot} t} \hat{\sigma}_z e^{-i \hat{H}_{\rm tot} t}
\nonumber \\ &=& \sum_{\alpha,\alpha'} \langle \alpha \vert
\hat{\rho}(0) \vert \alpha' \rangle \langle \alpha' \vert
\hat{\sigma}_z \vert \alpha \rangle e^{i (E_{\alpha}-E_{\alpha'})
t}. \label{Ast}
\end{eqnarray}

The time-averaged value of $\langle \hat{\sigma}_z \rangle(t)$ is
obtained performing the following time average:
\begin{eqnarray}
\langle \hat{\sigma}_z \rangle_{\infty}&=&\lim_{T\to\infty}
\frac{1}{T} \int_{0}^{T} dt \langle \hat{\sigma}_z \rangle(t)
\nonumber \\ &=&\sum_{\alpha} \langle \alpha \vert \hat{\rho}(0)
\vert \alpha \rangle \langle \alpha \vert \hat{\sigma}_z \vert
\alpha \rangle. \label{Asinf}
\end{eqnarray}
We see that this time averaged value clearly depends on the
initial condition.

An important result, that holds everywhere on the
parameter space and for all kinds of initial conditions, is
that the observed equilibrium value given by the exact von Neumann
equation corresponds very well to the time-averaged value (\ref{Asinf}).
Therefore, the study of the equilibrium properties of systems as ours
is equivalent to studying the time averaged quantities (\ref{Asinf}).

Let us note that the standard initial condition we used till
now is a microcanonical distribution around energy $\epsilon$ for
the environment and an upper state for the system formally given by
\begin{eqnarray}
\hat{\rho}(0) = \vert 1 \rangle \langle 1 \vert \otimes
\frac{\delta(\hat{H}_{B}-\epsilon)} {n(\epsilon)} = \sum_n
\frac{\delta(E_n-\epsilon)} {n(\epsilon)} \vert 1 n \rangle
\langle 1 n \vert. \label{rhozerostand}
\end{eqnarray}
Therefore, we have that
\begin{eqnarray}
\langle \alpha \vert \hat{\rho}(0) \vert \alpha \rangle = \sum_n
\frac{\delta(E_n-\epsilon)} {n(\epsilon)} \vert \langle \alpha
\vert 1 n \rangle \vert^2. \label{arhoamiccan}
\end{eqnarray}

\begin{figure}[h]
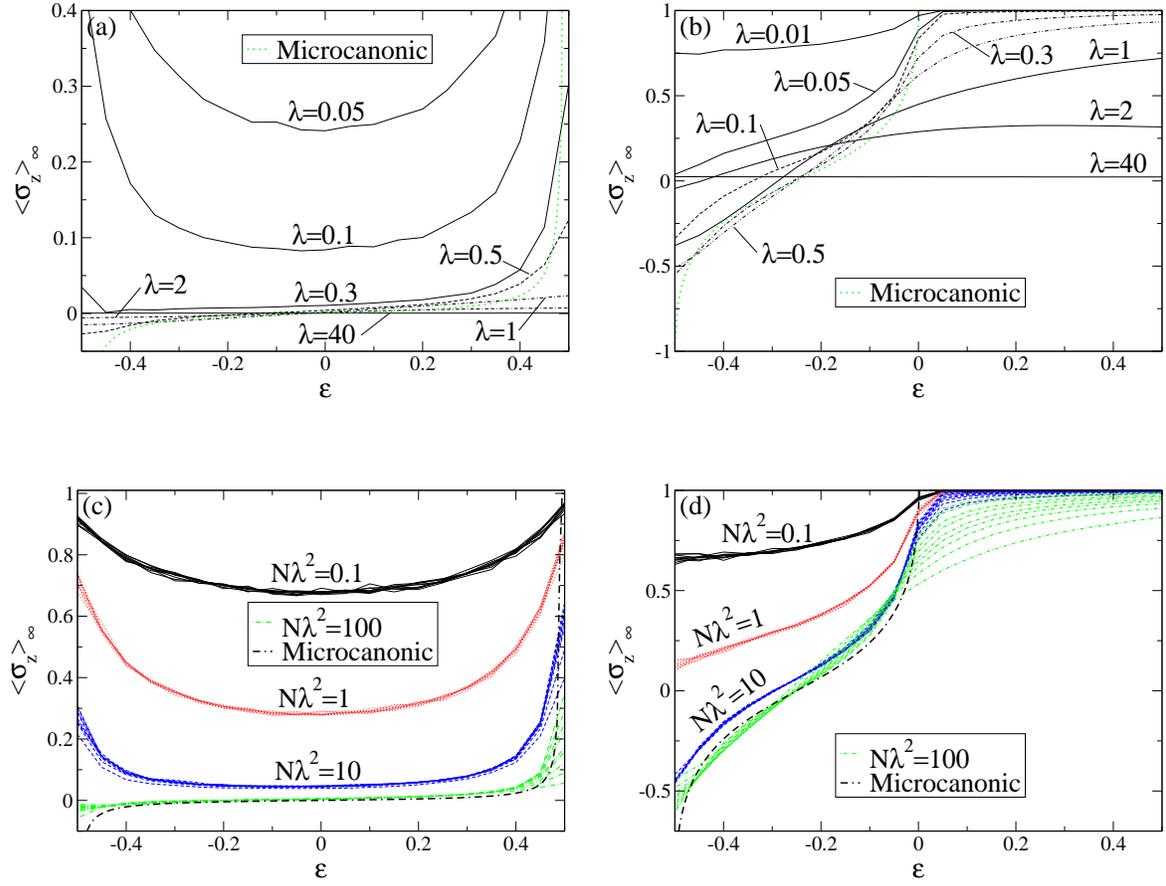

\centering
\begin{tabular}{c@{\hspace{0.5cm}}c}
\vspace*{1cm}
\rotatebox{0}{\scalebox{0.3}{\includegraphics{fig39.eps}}}
&
\rotatebox{0}{\scalebox{0.3}{\includegraphics{fig40.eps}}}
\\
\rotatebox{0}{\scalebox{0.3}{\includegraphics{fig41.eps}}}
&
\rotatebox{0}{\scalebox{0.3}{\includegraphics{fig42.eps}}}
\\
\end{tabular}
\caption{Comparison between between the equilibrium value of
$\langle \hat{\sigma}_z \rangle_{\infty}$ given by the Pauli equation and
the exact values given by the time averaged value for different
values of $\lambda$. These equilibrium values are depicted as a function
of the initial microcanonical energy $\epsilon$ of the environment. 
$\delta \epsilon=0.05$ in all figures. The parameter values are 
(a) $\Delta=0.01$ and $N=500$; (b) $\Delta=0.5$ and $N=500$; 
(c) $\Delta=0.01$ and $N=200-2000$; (d) $\Delta=0.5$ and $N=200-2000$.}
\label{sigZinftavvspauli}
\end{figure}

An interesting point is to understand when the Markovian
perturbative equation gives the correct equilibrium value. We
want, therefore, to compare (\ref{Asinf}) with
(\ref{paulispingoeMZinfini}). This is done in
Fig. \ref{sigZinftavvspauli} where we plotted the time
averaged value $\langle \hat{\sigma}_z \rangle_{\infty}$ as
function of the initial energy of the environment $\epsilon$ (the
initial value of the total system being (\ref{rhozerostand})). The
different curves in Figs. \ref{sigZinftavvspauli}(a) and 
\ref{sigZinftavvspauli}(b)
correspond to different values of $\lambda$. 
We compare these curves to the $\lambda$ independent
curves given by the Markovian perturbative equation
(\ref{paulispingoeMZinfini}).  As expected from our precedent
study of the dynamics, we find again that, when $\lambda$ is too
small, the "mixing" between the nonperturbed levels is not sufficient 
inside the microcanonical energy shell and the Markovian perturbative
results overestimate the equilibrium values. If $\lambda$ is too
large, the Markovian perturbative equation gives again bad
results. The Markovian perturbative equilibrium values are correct
in a characteristic region of $\lambda$. The beginning of this
region corresponds, in the SOE, to the critical
value of $\lambda$ at which the transition occurs from the localized to
the Lorentzian regimes. The end of this region corresponds to
$\lambda$ values that cannot be considered any more as
perturbative. Figures \ref{sigZinftavvspauli}(c) and 
\ref{sigZinftavvspauli}(d)
show that $\langle \hat{\sigma}_z \rangle_{\infty}$
scales like $\lambda^2 N$. This again confirms our precedent
analysis.

\subsection{Thermalization of the system}

One of the important questions is to understand the conditions
under which the system thermalizes under the effect of a weak contact with the
environment or in other words, under which conditions the
system relaxes to a canonical distribution corresponding to the
microcanonical temperature of the environment.

We begin by recalling these conditions in the general case of
a small system weakly interacting with its environment. The
isolated total system is composed of the system and the environment
and has the total energy
\begin{eqnarray}
E_{\rm tot} = \epsilon + e.
\end{eqnarray}
$\epsilon$ is the energy of the environment and $e$ the energy of
the system. We suppose that the contribution of the interaction
energy between the environment and the system is negligible
compared to the total energy. The microcanonical environment
entropy is defined as
\begin{eqnarray}
S_B(\epsilon)= k \ln \Omega_B(\epsilon), \label{Srho}
\end{eqnarray}
where $\Omega_B(\epsilon)$ is the number of states of the
environment available at energy $\epsilon$. This number can be
related to the density of states of the environment
$n_B(\epsilon)$ using the fact that $\Omega_B(\epsilon) =
n_B(\epsilon) \delta\epsilon$, where $\delta\epsilon$ is a small
energy interval but contains many states of the environment.
The microcanonical temperature of the environment is given by
\begin{eqnarray}
\frac{1}{T_B(\epsilon)} &=& \frac{d S_B(\epsilon)}{d \epsilon}  \label{defTmic}.
\end{eqnarray}
It can be expanded in the system energy as
\begin{eqnarray}
T_B(\epsilon=E_{\rm tot}-e) &=& T_B(\epsilon=E_{\rm tot}) - e \frac{d
T_B(\epsilon=E_{\rm tot})}{d \epsilon} + \dots ,
\end{eqnarray}
because we suppose that the system energy is much smaller than the
environment energy. The specific heat capacity of the environment
is
\begin{eqnarray}
\frac{1}{C_{vB}(\epsilon)} = \frac{d T_B(\epsilon)}{d \epsilon} .
\end{eqnarray}
If the condition
\begin{eqnarray}
\vert C_{vB}(\epsilon) \vert \gg \Big\vert \frac{e}{T_B(\epsilon)} \Big\vert \label{condenv2}
\end{eqnarray}
is satisfied, the temperature expansion can be truncated as
follows
\begin{eqnarray}
T_B(\epsilon=E_{\rm tot}-e) = T_B(\epsilon=E_{\rm tot}) \label{egaltemp}.
\end{eqnarray}
Therefore, we understand that if Eq. (\ref{condenv2}) is satisfied,
the environment plays the role of a heat bath because its temperature is
almost not affected by the system energy.

We suppose further that the interaction between the system and the
environment, even if small, is able to make the total probability
distribution microcanonical on the total energy shell at energy $E_{\rm tot}$.
We suppose also that the energy levels are discrete and, therefore, that
$E_{\rm tot}=E_s+E_b$ where $s$ and $b$ are discrete index's,
respectively, for the system and the environment. Therefore, the
probability $P_S(E_s)$ for the system being at energy $E_s$ is
given by
\begin{eqnarray}
P_S(E_s)=\frac{\Omega_B(E_b=E_{\rm tot}-E_s)}{\Omega_{\rm tot}(E_{\rm tot})},
\end{eqnarray}
where $\Omega_B(E_b)$ is the number of states of the environment
available at energy $E_b$, and $\Omega_{\rm tot}(E_{\rm tot})$ the number
of states of the total system available at energy $E_{\rm tot}$.
Using Eqs. \ref{defTmic}) and (\ref{condenv2}), one gets that
\begin{eqnarray}
P_S(E_s)=\frac{e^{\frac{1}{k} S_B(E_b=E_{\rm tot}-E_s)}}{\Omega_{\rm tot}(E_{\rm tot})} \simeq
\frac{ e^{ \frac{1}{k}S_B(E_b=E_{\rm tot})-\frac{E_s}{kT_B(E_B=E_{\rm tot})}} }
{\Omega_{\rm tot}(E_{\rm tot})} .
\end{eqnarray}
Using the normalization $ P_S(E_s)=1$, one finally gets the well-known canonical probability distribution for the
system
\begin{eqnarray}
P_S(E_s)=\frac{e^{-\frac{E_s}{k T_B(E_B=E_{\rm tot})}}}{Z},
\end{eqnarray}
where $Z=\sum_s e^{-\frac{E_s}{k T_B(E_B=E_{\rm tot})}}$.\\

We conclude that two conditions are necessary in order to thermalize the
system to a canonical probability distribution due to the
contact with the environment: a large heat capacity of the
environment $\vert C_{vB}(\epsilon) \vert \gg \vert
\frac{e}{T_B(\epsilon)} \vert$ and a microcanonical distribution
on the total energy shell.

\begin{figure}[h]
\centering
\rotatebox{0}{\scalebox{0.4}{\includegraphics{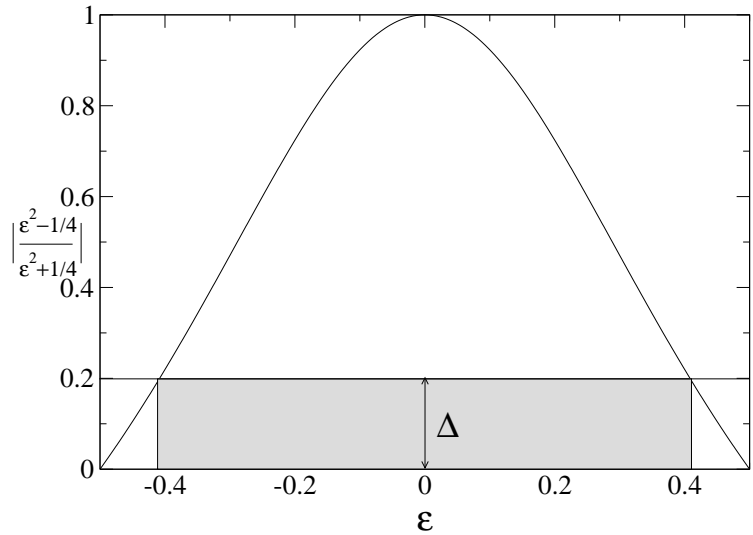}}}
\caption{Representation of the region where the thermalization condition $\Big\vert
\frac{\epsilon^2-\frac{1}{4}}{\epsilon^2+\frac{1}{4}} 
\Big\vert > \Delta$ holds.} \label{CvT}
\end{figure}

\begin{figure}[h]
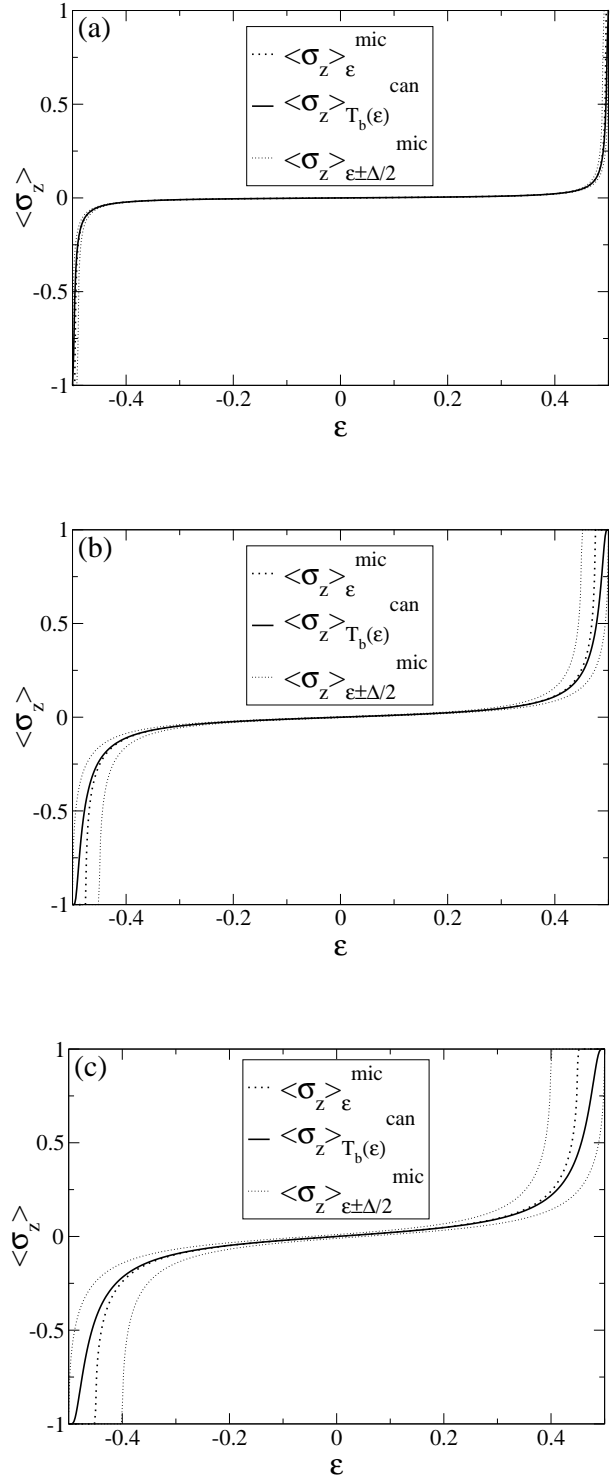

\centering
\begin{tabular}{c@{\hspace{0.5cm}}c}
\vspace*{1cm}
\rotatebox{0}{\scalebox{0.33}{\includegraphics{fig44.eps}}} \\
\vspace*{1cm}
\rotatebox{0}{\scalebox{0.33}{\includegraphics{fig45.eps}}} \\
\rotatebox{0}{\scalebox{0.33}{\includegraphics{fig46.eps}}}
\end{tabular}
\caption{Comparison between Eqs. (\ref{sigZcan}) and (\ref{sigZmic})
for different values of $\Delta$. The narrow dotted lines are plotted
to show the undetermination around the environment energy.
The parameter values are 
(a) $\Delta=0.01$; (b) $\Delta=0.05$; (c) $\Delta=0.1$.
"mic" means microcanonic and "can" means canonic.} 
\label{miccancomp}
\end{figure}

Let us apply this result to the spin-GORM model.
The density of states of the environment is
\begin{eqnarray}
n_B(\epsilon)=\frac{4N}{\pi} \sqrt{\frac{1}{4}-\epsilon^2}.
\end{eqnarray}
Therefore the microcanonical temperature of the environment is
\begin{eqnarray}
T_B(\epsilon)=\frac{\epsilon^2-\frac{1}{4}}{k \epsilon}
\end{eqnarray}
and the heat capacity is
\begin{eqnarray}
C_{vB}(\epsilon)=\frac{k \epsilon}{\epsilon^2+\frac{1}{4}}.
\end{eqnarray}
The canonical distribution of the populations of the system, at the
microcanonical temperature of the environment, is given by
\begin{eqnarray}
\langle \hat{\sigma}_z \rangle^{\rm can}_{T_B(\epsilon)} = - {\rm tanh}\frac{\Delta}{2 k T_B(\epsilon)}
= - {\rm tanh}\frac{\Delta\epsilon}{2 (\epsilon^2-\frac{1}{4})} . \label{sigZcan}
\end{eqnarray}
The first condition (\ref{condenv2}) in order to thermalize 
the system to a canonical probability distribution becomes
\begin{eqnarray}
\vert C_{vB}(\epsilon) T_B(\epsilon) \vert = \Big\vert
\frac{\epsilon^2-\frac{1}{4}}{\epsilon^2+\frac{1}{4}} \Big\vert \gg
\Delta \label{condenv3}
\end{eqnarray}
and is depicted in Fig \ref{CvT}.
The second condition to have a microcanonical distribution on the
total energy shell is satisfied (as we discussed in Sec.
\ref{valequiobssys}) when $\lambda^2 N > 1$. In this case, the
populations of the system obey the microcanonical equilibrium
value of the Markovian version of our perturbative equation
(\ref{paulispingoeMZinfini}), i.e.,
\begin{eqnarray}
\langle \hat{\sigma}_z \rangle^{\rm micro}_{\epsilon} =
\frac{\sqrt{\frac{1}{4}-(\epsilon-\frac{\Delta}{2})^2}-
\sqrt{\frac{1}{4}-(\epsilon+\frac{\Delta}{2})^2} }
{\sqrt{\frac{1}{4}-(\epsilon-\frac{\Delta}{2})^2}
+\sqrt{\frac{1}{4}-(\epsilon+\frac{\Delta}{2})^2}}.
\label{sigZmic}
\end{eqnarray}
If the two conditions are satisfied, then Eqs. (\ref{sigZcan}) and (\ref{sigZmic})
should be equal. The comparison between Eqs.
(\ref{sigZcan}) and (\ref{sigZmic}) can be seen in Fig. \ref{miccancomp} for
different system energies. We see that the smaller the system
energy is the better the comparison is. 

Therefore, we can conclude that under these two conditions ($\vert
\frac{\epsilon^2-\frac{1}{4}}{\epsilon^2+\frac{1}{4}} \vert \gg
\Delta$ and $\lambda^2 N > 1$), the random matrices of the
spin-GORM model can model an environment that behaves as a heat
bath.

\subsection{Thermalization of the total system}

Until now, we have chosen initial conditions where the system is
in the upper state with a microcanonical environment at a given
energy, like in Eqs. (\ref{rhozerostand}) or (\ref{arhoamiccan}). We now
want to consider initial conditions where the system is
again in the upper state but where the environment is at a given
canonical temperature:
\begin{eqnarray}
\hat{\rho}(0) = \vert 1 \rangle \langle 1 \vert \otimes
\frac{e^{-\beta_{B} \hat{H}_{B}}}{Z_{B}} = \sum_n
\frac{e^{-\beta_{B} E_{n}}}{Z_{B}} \vert 1 n \rangle \langle 1 n
\vert.
\end{eqnarray}
Therefore
\begin{eqnarray}
\langle \alpha \vert \hat{\rho}(0) \vert \alpha \rangle = \sum_n
\frac{e^{-\beta_{B} E_{n}}}{Z_{B}} \vert \langle \alpha \vert 1 n
\rangle \vert^2. \label{arhoacan}
\end{eqnarray}
It is important to notice that there is no statistical equivalence
between the canonical and the microcanonical ensembles in the spin-GORM
model (see Appendix B). Therefore, it is interesting to ask how
the probability distribution looks like at equilibrium after the
interaction.

One can clarify this point by plotting $\langle \alpha \vert \hat{\rho}(0)
\vert \alpha \rangle$ versus energy. One uses the following energy representation
\begin{eqnarray}
P(\varepsilon)= \sum_{\alpha} \delta(E_{\alpha}-\varepsilon)
\langle \alpha \vert \hat{\rho}(0) \vert \alpha \rangle.
\end{eqnarray}

If the total system thermalizes and reaches a canonical distribution
for the total system at an effective temperature
$\beta_{\rm eff}^{-1}$, one would have that
\begin{eqnarray}
P(\varepsilon)= \frac{e^{-\beta_{\rm eff} \varepsilon}}{Z_{\rm tot}},
\end{eqnarray}
because
\begin{eqnarray}
\langle \alpha \vert \hat{\rho}(0) \vert \alpha \rangle =
\frac{e^{-\beta_{\rm eff} E_{\alpha}}}{Z_{\rm tot}}.
\label{TOTcanonicavAs}
\end{eqnarray}

As we shall see, it is the case if, again, $\lambda$ is large
enough to induce "mixing" between the states $\lambda^2 N > 1$.

\begin{figure}[h]
\centering
\rotatebox{0}{\scalebox{0.3}{\includegraphics{fig47.eps}}} \\
\centering
\vspace*{1cm}
\begin{tabular}{c@{\hspace{0.5cm}}c}
\rotatebox{0}{\scalebox{0.3}{\includegraphics{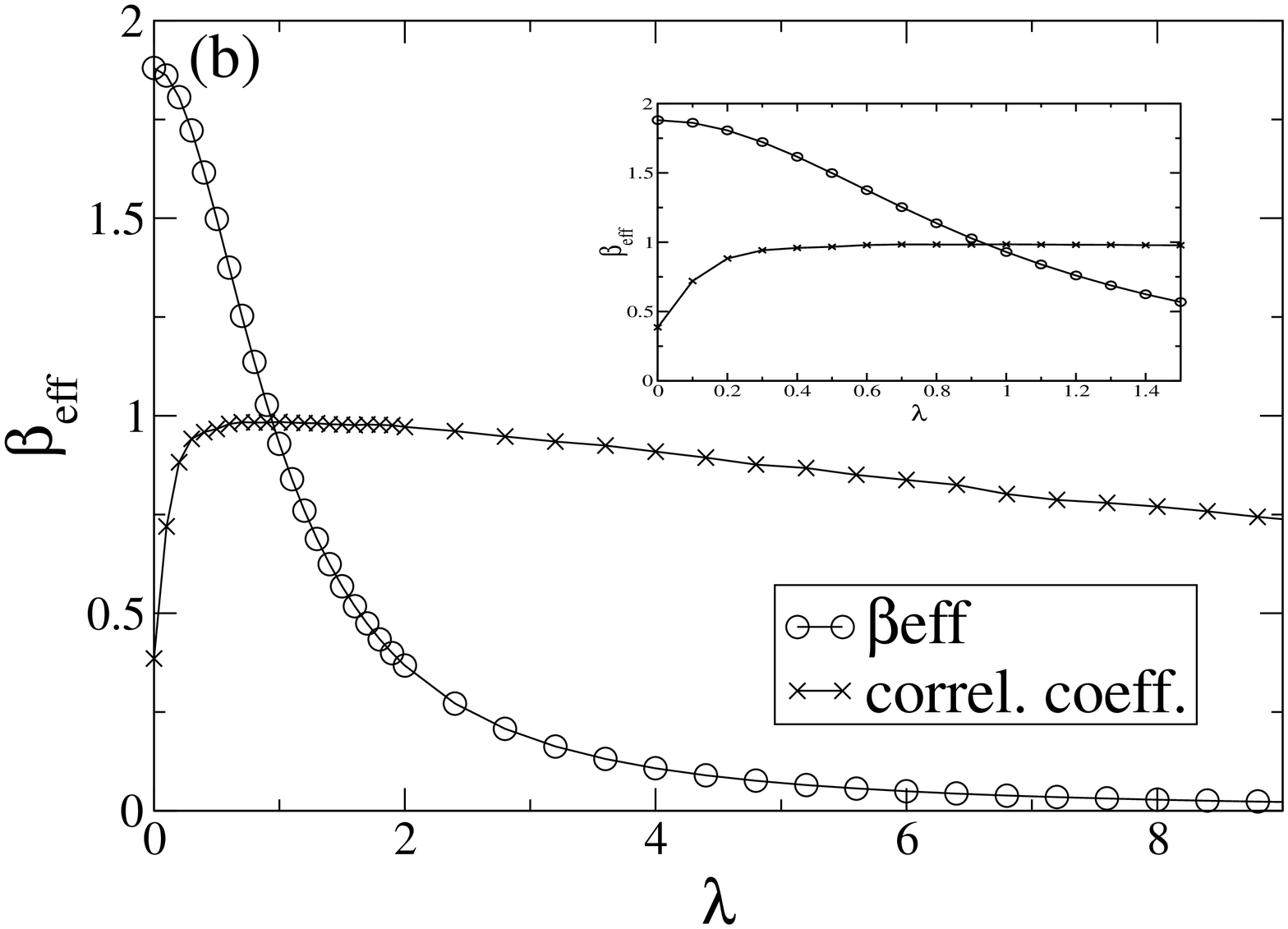}}}
&
\rotatebox{0}{\scalebox{0.3}{\includegraphics{fig49.eps}}}
\\
\end{tabular}
\caption{In all the figures $\Delta=0.01$, $N=500$, and $\beta=2$. 
(a) Probability $P(\varepsilon)={\rm Dist}(\varepsilon)$ 
of being in an eigenstate of 
the total system at equilibrium starting from the initial condition 
$\hat{\rho}(0) = \vert 1 \rangle \langle 1 \vert 
\frac{e^{-\beta \hat{H}_B}}{Z_B}$
with $\frac{1}{\beta} = 0.5$. (b) Effective temperature of the equilibrium probability
distribution obtained by fitting a canonical distribution to the data of (a). 
(c) Comparison between the exact equilibrium population value and the canonical one
computed with the effective temperature of plot (b).}
\label{thermaB}
\end{figure}

Indeed, one sees in Fig. \ref{thermaB}(a) that, for $\lambda=0$,
the states of the total system corresponding to the upper level of
the system are exponentially populated and the ones corresponding
to the lower level are not. When the interaction is turned on and
increased, one can notice that the probability distribution starts
to accumulate around a mean effective canonical distribution. As
expected, this accumulation becomes significant when $\lambda^2 N >
1$ and, in this case, the total system can be considered as having
thermalized. One can calculate the final effective temperature
that the total system has reached after interaction. This
effective temperature is depicted in
Fig. \ref{thermaB}(b) as a function of $\lambda$. The correlation
coefficient indicates whether the exponential fit of the
final effective temperature is good or not. The effective
temperature obeys the following law:
$\beta_{eff}=\frac{\beta_{i}}{1+\lambda^2}$. We also show in
Fig. \ref{thermaB}(c) the comparison between the time-averaged
value of $\langle \hat{\sigma}_z \rangle$ and his canonical average
computed with the effective temperature. One sees that, when
$\lambda^2 N > 1$, both coincide.

This thermalization is not statistical in the sense of the
equivalence between the ensembles. It is an intrinsic
thermalization due to the complexity of the interaction between
the states. This thermalization appears at a critical value of the
coupling parameter when the interaction term becomes of the order
of or larger than the mean level spacing of the total system. 

\section{Conclusions}
\label{conclusions}

In this paper, we have studied a system made of two parts: a two-level system
interacting in a nondiagonal way with a complex environment
modeled by Gaussian orthogonal random matrices.

We began our study by analyzing the spectral properties of this
model. We investigated the spectrum on a large energy scale with
the averaged smooth density of states and on a finer energy scale
with the eigenvalue diagrams, the shape of the eigenstates
(SOE), the spacing distribution, and the asymptotic transition
probability kernel (ATPK). We found a global repulsion 
as well as avoided crossings between the
eigenvalues when the coupling parameter $\lambda$ was increased. 
We also showed the existence of three regimes (easy to distinguish in the SOE) that
are important to describe the different qualitative behaviors
of the model: the localized regime when the interaction between
the levels is weaker then the mean level spacing $\lambda^2 N
\lesssim 1$ (giving rise to very narrow SOE), the Lorentzian
regime when the interaction between the levels becomes larger then
the mean level spacing $\lambda^2 N > 1$ (giving rise to a
Lorentzian SOE), and the delocalized regime for very large $\lambda$
(giving rise to SOE spread over the whole spectrum).

\begin{figure}[h]
\centering
\rotatebox{0}{\scalebox{0.7}{\includegraphics{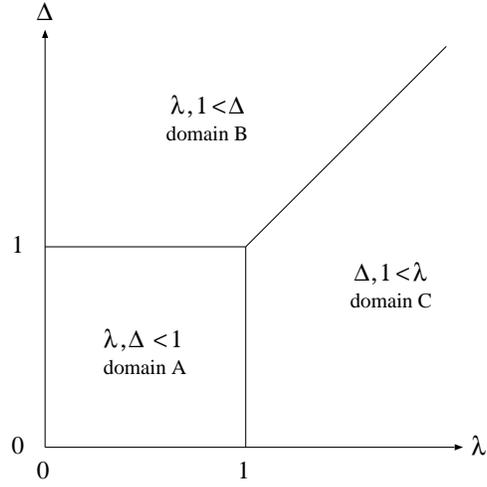}}} \\
\caption{Representation of the three different domains
in the space of the reduced parameters $\lambda$ and $\Delta$
of the model for a fixed number $N$ of states.}
\label{regimeschema}
\end{figure}

After the spectral study, we started the study of the dynamics of
the system populations. We defined different domains in the
parameter space (see Fig. \ref{regimeschema}) and related
them to the different relaxation behaviors of the system
population induced by the interaction with the environment. For
each of these domains we tested the validity of approximated
population evolution equations. In the strong coupling limit, we
identified a population relaxation regime described by a Bessel
function: $\sim 2\frac{J_1 (\lambda t)}{\lambda t}$ (\ref{Besstlsurtl}) 
that scales in
time according to $\lambda t$ and reaches an equilibrium distribution
corresponding to the same probability of being in the 
upper and lower states of the system. In the small coupling
limit and for small system energy, we obtained a Pauli-type equation
(\ref{paulispingoeMZt})-(\ref{paulispingoeMrate}) describing an 
exponential relaxation of the system population that
scales in time according to $\lambda^2 t$ and that reaches an
equilibrium distribution depending on the system energy. The
equilibrium value corresponds to a microcanonical probability
distribution of being in a nonperturbed state of the total system
inside the total energy shell. Finally, we showed the necessity of
taking into account the non-Markovian effects in the dynamics
(which are important when the system energy becomes non-negligible
in front of the environment energy) using a non-Markovian
perturbative equation (\ref{zdefpauligen1}) and (\ref{paulispingoeNM11})
derived by the authors in Ref. \cite{Esposito}.
This equation is perturbative and therefore only valid for small
coupling parameters. This equation describes the highly
non-Markovian dynamics of the population (made of small and fast
system oscillations damped on a time scale corresponding to the
environment time scale) when the system energy becomes large in
front of the environment energy. This equation also reduces to the
Pauli-type equation in the opposite situation, when the system
energy is small compared to the environment energy.  The validity of
these approximated equations depend on the parameter domain
considered and are summarized in Fig. \ref{valideq}.  An
important result concerning the small coupling limit is that
there exist lower and upper bounds on the coupling
parameter values for which the pertubative equation holds.
The lower bound depends on the spacing between the states of the
total systems and therefore on the size $N$ of the random matrices
modeling the environment. We showed that this lower bound is
related to the transition between the localized and the Lorentzian
regimes in the SOE and that this bound disappears when $N \to
\infty$.

Another important result concerns the equilibrium values of the
spin-GORM model. We showed that they are very well reproduced by
the time averaged quantities. 
Moreover, we showed that the spin-GORM model can, under two conditions,
describe the thermalization of the system to a canonical energy
probability distribution corresponding to the environment
microcanonical temperature. The two conditions for this
thermalization are: a microcanonical probability distribution on the
energy shell of the total system (that occurs when $\lambda^2 N
> 1$) and a large heat capacity of the environment compared to the
ratio between the characteristic system energy and the environment
temperature.

Finally, we showed that the spin-GORM model can undergo an
intrinsic thermalization due to the complex interaction between
the states, and reach an overall thermal canonical distribution. 
This thermalization again occurs when the coupling
parameter is large enough (i.e., larger than the mean level spacing of
the total system) to "mix" the levels.

\newpage
\centerline{\bf APPENDIX A: Gaussian orthogonal random matrices
(GORM)}
\vspace{1.5cm}

A Gaussian orthogonal random matrix (GORM) $\hat{Y}$ is
characterized by $M$, the size of the matrix, and by the parameter
$a_{\hat{Y}}$, which enters the Gaussian probability distribution
$P(\hat{Y})=C e^{- \frac{a_{\hat{Y}}}{2} Tr(\hat{Y}^2)}$ of the
whole matrix. The statistical properties of a GORM are preserved
under orthogonal transformations. Because the matrix is
orthogonal, each nondiagonal element $Y_{ij}$ is equal to its
transposed $Y_{ji}$. The $\frac{M(M+1)}{2}$ independent matrix
elements of $\hat{Y}$ are Gaussian random numbers of mean zero.
The standard deviation of the nondiagonal matrix elements
$\sigma_{ND}^{\hat{Y}}$ and the standard deviation of the diagonal
matrix element $\sigma_{D}^{\hat{Y}}$ are related to $a_{\hat{Y}}$
by
\begin{equation}
\sigma_{D}^{\hat{Y}}=\sqrt{2}
\sigma_{ND}^{\hat{Y}}=\sqrt{\frac{1}{a_{\hat{Y}}}}.
\end{equation}
The \textit{density of states} of the GORM $\hat{Y}$ is defined by
\begin{equation}
d(E)=\sum_{i=1}^M \delta(E-E_i),
\end{equation}
and the \textit{smoothed density of states} by
\begin{equation}
\bar{d}(E)= \lim_{\epsilon \to 0} \frac{1}{\epsilon}
\int_{E-\frac{\epsilon}{2}}^{E+\frac{\epsilon}{2}} d(E) \ \ dE,
\end{equation}
where $\epsilon$ is a small energy interval which is large enough to
contain many states in order for $\bar{d}(E)$ to be smooth. The
\textit{averaged smoothed density of states} is an ensemble
average of $\chi$ realizations of the GORM. Such an ensemble is
called the \textit{Gaussian orthogonal ensemble} (GOE). It is well
known \cite{Mehta,Porter,Brody} that the ensemble averaged smoothed density of
states $\langle \bar{d}(E) \rangle_{\chi}$ obey the
\textit{Wigner semicircle law} in the limit $\chi \to \infty$:
\begin{eqnarray}
\langle \bar{d}(E) \rangle_{\infty} &=& \frac{a_{\hat{Y}}}{\pi}
\sqrt{\frac{2M}{a_{\hat{Y}}}-E^2} \ \ if \ \ \vert E \vert <
\sqrt{\frac{2M}{a_{\hat{Y}}}} \nonumber \\ &=& 0 \ \ if \ \ \vert
E \vert \geq \sqrt{\frac{2M}{a_{\hat{Y}}}}. \label{semi-circ}
\end{eqnarray}
The domain of energy where the eigenvalues are distributed (i.e., the
width of the semi-circle) is
$\mathcal{D}\!Y=\sqrt{\frac{8M}{a_{\hat{Y}}}}$. Notice that when
$M \to \infty$, $\bar{d}(E) \to \langle \bar{d}(E)
\rangle_{\infty}$, and therefore $\bar{d}(E)$ follows the
semi-circle law. The following notation is used in the present
paper: $n^w(E)=\langle \bar{d}(E) \rangle_{\infty}$.

\newpage
\centerline{\bf APPENDIX B: Equivalence between ensembles for the environment}
\vspace{1.5cm}

Let us consider a system interacting with its environment.
The environment is in a canonical distribution at temperature $T_{\rm can}=\frac{1}{k \beta}$.
The evolution of a system observable is given by
\begin{eqnarray}
\langle \hat{A}_S \rangle^{\beta}(t)&=&\textrm{Tr} \ \
\hat{\rho}_S(0) \frac{e^{-\beta\hat{H}_B}}{Z_B} e^{i
\hat{H}_{\rm tot} t} \hat{A}_S e^{-i \hat{H}_{\rm tot} t}. \label{evolAscan}
\end{eqnarray}
On the other hand, if the environment is in a microcanonical ensemble, the
evolution of the system observable is given by
\begin{eqnarray}
\langle \hat{A}_S \rangle^{\epsilon}(t)&=&\textrm{Tr} \ \
\hat{\rho}_S(0) 
\frac{\delta(\epsilon-\hat{H}_B)}{n(\hat{H}_B )} e^{i
\hat{H}_{\rm tot} t} \hat{A}_S e^{-i \hat{H}_{\rm tot} t}. \label{evolAsmiccan}
\end{eqnarray}
One therefore sees that
\begin{eqnarray}
\langle \hat{A}_S \rangle^{\beta}(t)&=& \int d\epsilon n(\epsilon)
\frac{e^{-\beta \epsilon}}{Z_B} \langle \hat{A}_S
\rangle^{\epsilon}(t). \label{equivexacte}
\end{eqnarray}
The statistical equivalence between the canonical and 
microcanonical ensembles, $\langle \hat{A}_S
\rangle^{\beta}(t)=\langle \hat{A}_S \rangle^{\epsilon'}(t)$,
thus occurs when
\begin{eqnarray}
n(\epsilon) \frac{e^{-\beta \epsilon}}{Z_B} \approx
\delta(\epsilon-\epsilon'). \label{equivcond}
\end{eqnarray}
One understands that this equivalence is qualitatively satisfied
when $n(\epsilon)$ is an incresing function of $\epsilon$. In this case,
$n(\epsilon) \frac{e^{-\beta \epsilon}}{Z_B}$ is a sharply peaked function.
In order to find the maximum of $n(\epsilon) \frac{e^{-\beta \epsilon}}{Z_B}$, we
require the vanishing of the derivative of its logarithm
$\frac{\partial}{\partial\epsilon} \ln \left[n(\epsilon)
\frac{e^{-\beta \epsilon}}{Z_B}\right]=0$. We find that
\begin{eqnarray}
\frac{\partial}{\partial\epsilon}
S(\epsilon_{\rm max})\equiv\frac{1}{T_{\rm micro}(\epsilon_{\rm max})}\approx\frac{1}{T_{\rm can}},
\label{entropmic}
\end{eqnarray}
where the microcanonical entropy is given by
\begin{eqnarray}
S(\epsilon)=k \ln n(\epsilon) \delta \epsilon,
\end{eqnarray}
where $\delta \epsilon$ is a small energy shell containing many levels.
This shows that if $n(\epsilon) \frac{e^{-\beta \epsilon}}{Z_B}$ is a sharply
peaked function around $\epsilon_{\rm max}$, the canonical average at temperature $T_{\rm can}$ is
equivalent to the microcanonical average at the energy $\epsilon_{\rm max}$ corresponding
to the microcanonical temperature $T_{\rm micro}(\epsilon)=T_{\rm can}$.

For the spin-GORM model, there is no equivalence between the
canonical and the microcanonical ensembles. It is due to the fact that the
semicircular energy distribution is not an increasing function of the energy.
In this sense, this shows that the semicircular energy
distribution does not describe a usual environment.

To complete our reasoning, we notice that the maximum of $n(\epsilon) \frac{e^{-\beta \epsilon}}{Z_B}$
is given by
\begin{eqnarray}
\epsilon_{\rm max}=\frac{1-\sqrt{1+\beta^2}}{2 \beta},
\end{eqnarray}
so that $\epsilon_{\rm max}\stackrel{\beta\to0}{=}0$ and 
$\epsilon_{\rm max}\stackrel{\beta\to\infty}{=}-\frac{1}{2}$.


\newpage

\centerline{\bf APPENDIX C: Perturbation theory} \vspace{1.5cm}

There is no analytical way of getting a general form of the
eigenvalues $E_{\alpha}$ of the total system, 
but the three terms in Eq. (\ref{hamiltonien}) have
different orders of magnitude depending of the value of the
parameters $\Delta$ and $\lambda$. The system and the environment
Hamiltonians are, respectively, of order $\Delta$ and $1$ while the
coupling term is of order $\lambda$. Therefore, we can examine the
different extreme cases that can be treated perturbatively.\\

\boldmath $\Delta,1 \gg \lambda$ \unboldmath :\\

When the system and the environment Hamiltonians are larger than
the interaction term in Eq. (\ref{hamiltonien}), we can treat the
interaction term in a perturbative way, taking the system and the
environment Hamiltonian as reference,
\begin{equation}
\hat{H}_0 \vert s b \rangle
= E_{s b}^{0} \vert s b \rangle \; ,
\end{equation}
where we replaced the index ${n}$ by the two indices ${s,b}$.
The perturbed energy is given to the second order by
\begin{equation}
E_{\alpha} = E_{s,b} = \frac{\Delta}{2} s + E_{b}^{B}
+ \lambda^2 \sum_{b' \neq b} \frac{\vert \langle b' \vert \hat{B}
\vert b \rangle \vert^2}{E_{b}^{B}-E_{b'}^{B}+ s \Delta} +O(\lambda^4).
\label{pertcas1}
\end{equation}
We notice that the first nonzero correction to the
nonperturbed eigenstate is of order $\lambda^2$.\\

\boldmath $\lambda \gg 1,\Delta$ \unboldmath :\\

When $\lambda$ is large compared to $\Delta$ and $1$ in Eq.
(\ref{hamiltonien}), it is possible to consider the interaction
term as the reference Hamiltonian and to treat $\hat{H}_S$ and
$\hat{H}_B$ as small perturbation. Transforming
(\ref{hamiltonien}) by a unitary matrix acting only on the system
degree of freedom, we get
\begin{eqnarray}
\hat{H}_{\rm tot} = \frac{\Delta}{2} \hat{\sigma}_{x} + \hat{H}_B +
\lambda \hat{\sigma}_z \hat{B}.
\end{eqnarray}
The nonperturbed reference Hamiltonian is, therefore,
\begin{eqnarray}
\hat{\tilde{H}}_{0} = \lambda \hat{\sigma}_z \hat{B}.
\end{eqnarray}
Let $E_{\kappa \eta}$ and $\vert \kappa \eta \rangle = \vert
\kappa \rangle \otimes \vert \eta \rangle$ be, respectively, the
eigenvalues and eigenvectors of $\hat{\tilde{H}}_0$:
\begin{equation}
\hat{\tilde{H}}_0 \vert \kappa \eta \rangle = \lambda \hat{B}
\sigma_z \vert \kappa \eta \rangle = \lambda E_{\kappa \eta} \vert
\kappa \eta \rangle = \lambda \kappa E_{\eta} \vert \kappa \eta
\rangle,
\end{equation}
where $\eta=1,...,\frac{N}{2}$ and $\kappa=\pm1$. The energy of
the perturbed Hamiltonian is thus given to the second order perturbation
in $\frac{1}{\lambda}$ by
\begin{equation}
\frac{E_{\alpha}}{\lambda} = \kappa E_{\eta} + \frac{1}{\lambda}
\langle \eta \vert \hat{H}_B \vert \eta \rangle +
\frac{1}{\lambda^2} \sum_{\stackrel{\kappa',\eta'}{\neq
\kappa,\eta}} \frac{\vert \frac{\Delta}{2} + \langle \eta \vert
\hat{H}_B \vert \eta \rangle \vert^2}{E^{0}_{\kappa \eta} -
E^{0}_{\kappa' \eta'}}+ O\left(\frac{1}{\lambda^3}\right).
\label{pertcas2}
\end{equation}

\boldmath $1 \gg \Delta,\lambda$ \unboldmath :\\

In this case, the bath Hamiltonian is large compared to the system Hamiltonian
and the interaction term so that they both can be considered as perturbations.
We get
\begin{equation}
E_{\alpha} = E_{s,b} = \frac{\Delta}{2} s + E_{b}^{B}
+ \lambda^2 \sum_{b' \neq b} \frac{\vert \langle b' \vert \hat{B}
\vert b \rangle \vert^2}{E_{b}^{B}-E_{b'}^{B}}
+ O(\Delta^2) + O(\lambda^2).
\label{pertcas3}
\end{equation}

\boldmath $\Delta \gg 1,\lambda$ \unboldmath :\\

We now suppose that the system Hamiltonian, taken as reference, is large compared
to the environment Hamiltonian and the interaction term, so that these last two terms can
be considered as perturbations.
We then get
\begin{equation}
E_{\alpha} = E_{s,b} = \frac{\Delta}{2} s + E_{b}^{B}
+ s \frac{\lambda^2}{\Delta} \sum_{b' \neq b} \vert \langle b' \vert \hat{B}
\vert b \rangle \vert^2 + O(1) + O(\lambda^2).
\label{pertcas4}
\end{equation}

Two more situations, \boldmath $1,\lambda \gg \Delta$ \unboldmath and
\boldmath $\Delta,\lambda \gg 1$ \unboldmath, could be considered but
cannot be treated perturbatively because no reference
basis exists in which $\hat{H}_B$ and $\hat{B}$ are simultaneously diagonal.

\newpage

\noindent{\bf Acknowledgments}

The authors thank Professor G. Nicolis for support and
encouragement in this research, as well as D. Cohen for several
very fruitful discussions during his visit to Brussels. M.E. also
wants to thank I. de Vega for her interesting comments on the
spin-GORM model. M. E. is supported by the Fond pour la formation
\`{a} la Recherche dans l'Industrie et dans l'Agriculture, and P.
G. by the National Fund for Scientific Research (F.~N.~R.~S.
Belgium).\\


\newpage

\begin{figure}[h]
\centering
\rotatebox{0}{\scalebox{1.5}{\includegraphics{fig32.eps}}} \\
\caption{Table giving the validity of the approximated
equation and of the time scaling for the different parameter domains
of the moodel. The parameter domains are
defined in Fig. \ref{regimeschema}. 
"M pert. eq." means the Markovian verion of the perturbative equation 
(\ref{paulispingoeMZt}), "NM pert. eq." refers to the Markovian version of 
the perturbative equation (\ref{zdefpauligen1}), and "Bessel" refers to the 
equation (\ref{Besstlsurtl}) obtained in the very strong coupling regime.} 
\label{valideq}
\end{figure}

\end{document}